\documentclass{pasa}

\usepackage[utf8]{inputenc}
\usepackage[english]{babel}
\usepackage{graphicx}	
\usepackage{amsmath}	
\usepackage{amssymb}	
\usepackage[separate-uncertainty = true,multi-part-units=single]{siunitx}
\usepackage[linesnumbered]{algorithm2e}
\usepackage{newtxtext,newtxmath}
\usepackage{nicefrac}
\usepackage{subcaption}
\usepackage{booktabs}
\usepackage{physics}

\usepackage{aas_macros}
\usepackage{hyperref} 
\hypersetup{colorlinks,citecolor=blue,linkcolor=blue,urlcolor=blue}
\hypersetup{draft}

\addto\extrasenglish{
    
}

\DeclareSIUnit\parsec{pc}
\DeclareSIUnit\gauss{G}
\DeclareSIUnit\h{h}
\DeclareSIUnit\jansky{Jy}
\DeclareSIUnit\beam{beam}
\DeclareSIUnit\deg{deg}
\DeclareSIUnit\solarmass{M\textsubscript{\(\odot\)}}

\title[FIGARO Simulation]{FIGARO Simulation: FIlaments \& GAlactic RadiO Simulation}

\author[Hodgson et al.]{Torrance Hodgson$^{1}$\thanks{torrance@pravic.xyz}, Franco Vazza$^{2,3,4}$, Melanie Johnston-Hollitt$^{1,5}$ and Benjamin McKinley$^{1,6}$
\affil{$^1$International Centre for Radio Astronomy Research (ICRAR), Curtin University, 1 Turner Ave, Bentley, 6102, WA, Australia}
\affil{$^2$Dipartimento di Fisica e Astronomia, Universit\a'a di Bologna, Via Gobetti 92/3, 40121, Bologna, Italy}
\affil{$^3$Hamburger Sternwarte, Gojenbergsweg 112, 21029 Hamburg, Germany}
\affil{$^4$INAF, Istituto di Radio Astronomia di Bologna, Via Gobetti 101, 40129 Bologna, Italy}
\affil{$^5$Curtin Institute for Computation, Curtin University, GPO Box U1987, Perth, 6845, WA, Australia}
\affil{$^6$ARC Centre of Excellence for All Sky Astrophysics in 3 Dimensions (ASTRO3D), Bentley, Australia}
}

\jid{PASA}
\doi{10.1017/pas.\the\year.xxx}
\jyear{\the\year}

\begin{document}

\begin{frontmatter}
\maketitle

\begin{abstract}
We produce the first low to mid frequency radio simulation that incorporates both traditional extragalactic radio sources as well as synchrotron cosmic web emission. The FIlaments \& GAlactic RadiO (FIGARO) simulation includes ten unique \SI{4x4}{\degree} fields, incorporating active galactic nucleii (AGNs), star forming galaxies (SFGs) and synchrotron cosmic web emission out to a redshift of $z = 0.8$ and over the frequency range \SIrange[range-units=single]{100}{1400}{\mega \hertz}. To do this, the simulation brings together a recent $100^3$~Mpc$^3$ magneto-hydrodynamic simulation \citep{Vazza2019}, calibrated to match observed radio relic population statistics, alongside updated `T-RECS' code for simulating extragalactic radio sources \citep{Bonaldi2019}. Uniquely, the AGNs and SFGs are populated and positioned in accordance with the underlying matter density of the cosmological simulation. In this way, the simulation provides an accurate understanding of the apparent morphology, angular scales, and brightness of the cosmic web as well as---crucially---the clustering properties of the cosmic web with respect to the embedded extragalactic radio population. We find that the synchrotron cosmic web does not closely trace the underlying mass distribution of the cosmic web, but is instead dominated by shocked shells of emission surrounding dark matter halos and resembles a large, undetected population of radio relics. We also show that, with accurate kernels, the cosmic web radio emission is clearly detectable by cross-correlation techniques and this signal is separable from the embedded extragalactic radio population. We offer the simulation as a public resource towards the development of techniques for detecting and measuring the synchrotron cosmic web.
\end{abstract}

\end{frontmatter}



\section{Introduction}

The term `cosmic web' has been used to evoke the distribution of matter in our Universe on the very largest of scales. In this model, perturbations shortly after inflation have been amplified by gravitational instability to drive a process of hierarchical structure formation: low density regions have evolved into giant voids, high density regions have seeded galaxies, groups and clusters, and these are connected by a network of low density filaments and sheets \citep[e.g.][]{Baugh2004}. Outside of galaxies and cluster cores, some 40--50\% of the Baryonic mass of the Universe is believed to trace this cosmic web structure, existing as a diffuse, highly-ionised, warm-hot intergalactic medium \citep[e.g.][]{Cen1999, Dave2001}.

Until recently, the existence of the WHIM and its distribution in a cosmic web was primarily inferred from simulations of Universe evolution. Tentative empirical results have begun to support this model \citep[e.g.][]{Eckert2015, Nicastro2017, deGraaff2019, Tanimura2019}. Most recently, \citet{Macquart2020} provided compelling evidence in support of this hypothesis by tracing the dispersion measure of a small collection of fast radio bursts, the origins of each having been traced to a known host galaxy. Despite the small sample size, this result has provided the strongest evidence yet for these missing baryons residing along the line of sight in the intracluster medium.

The \textit{synchrotron} cosmic web is the expected radio component emitted by this large scale structure \citep{Brown2011}. As part of the ongoing process of large scale structure formation, simulations such as \citet{Vazza2015,Vazza2019} have modelled large-scale accretion processes and shown them capable of producing shocks surrounding filaments and the outermost regions of galaxy clusters many times the local speed of sound, with Mach numbers as high as $\mathcal{M} \sim$~\SIrange[range-phrase=--,range-units=single]{10}{100}{}. Such shocks are capable of producing high energy electrons by way of diffusive shock acceleration (DSA) and, in the presence of large-scale intracluster magnetic fields, this suprathermal population will in turn produce synchrotron emission \citep[e.g.][]{Keshet2009}. The strength of this emission is predicted to be extremely weak, however, and in previous modelling was predicted to be at or below the level of detectability of the current generation of low frequency radio instruments \citep{Vazza2015}. Using more recent cosmological numerical simulations as guidance, \citet{Gheller2020} tested the detectability of a number observables that trace the gas in the cosmic web using a cross-correlation technique. They reported that the observables with the strongest cross-correlation with the underlying distribution of galaxies should be those involving the Sunyaev Zeldovich effect and, in spite of the difficulty of its detection, the diffuse synchrotron radio emission from the shocked cosmic web.

Two separate papers by \citet{Brown2017} and \citet{Vernstrom2017} both attempted a statistical detection using a cross-correlation analysis. \citet{Vernstrom2017} used a deep, \SI{180}{\mega \hertz} observation of a \SI{21.8 x 21.8}{\degree} field using the Murchison Widefield Array \citep[MWA;][]{Tingay2013} and cross-correlated this (residual) map with galaxy density maps at various redshifts, smoothed to scales ranging from \SIrange[range-phrase=--,range-units=single]{1}{4}{\mega \parsec}. Their expectation was that there would be a peak in the cross-correlation at \SI{0}{\degree} offset, and this expectation was rooted in an assumption that cosmic web emission broadly traced the large-scale mass distribution of the Universe. However, other radio sources such as active galactic nuclei (AGNs) and star-forming galaxies (SFGs) also correlate with galaxy density maps, making it necessary to accurately model these related emission populations to distinguish their signals. Indeed, despite the authors detecting a peak in the correlation at \SI{0}{\degree} offset, it was this confounding factor that prohibited any claim to a positive detection.

\citet{Brown2017} similarly attempted to use a cross-correlation analysis using the S-Band Polarization All-Sky Survey observed with Parkes at \SI{2.3}{\giga \hertz} \citep{Carretti2013}. However, rather than using galaxy density as a proxy for the cosmic web, they used cosmological simulations that were constrained to reproduce the local large-scale structure and which tracked the evolution of thermal gas and magnetic field strengths. The resulting synchrotron cosmic web emission $S$ was modelled as a function of thermal electron density $n_e$ and magnetic field strength $B$ in the form $S \propto n_e B^2$. From this they produced a large-scale, low resolution map of the local synchrotron cosmic web showing that it broadly and smoothly traced out the underlying mass density of the local Universe. The cross-correlation showed no statistically significant detection.

Both papers point to the great difficulty of making a detection of the synchrotron cosmic web. In particular, they point to the need for future detection attempts to be able to accurately model how the radio emission traces the underlying matter density, as well as to understand the much brighter population of AGN and SFG radio sources, how these cluster with respect to the underlying synchrotron cosmic web, and how their emission may produce confounding signals.

Most recently, \citet{Vernstrom2021} have claimed the first definitive detection of the synchrotron cosmic web. By using luminous red galaxies as tracers of cluster cores, Vernstrom et al.\@ stacked nearby cluster pairs found in low frequency radio data produced both by the MWA as well as Owens Valley Radio Observatory Long Wavelength Array \citep[OVRO-LWA][]{Eastwood2018} and identified a residual signal produced in the spanning intracluster medium. The authors argue against alternative explanations for the signal such as intervening cluster emission or overdense AGN and SFG emission spanning the filaments; whilst the results of this experiment are promising, the work also points to the need to accurately understand and model the combined extragalactic population alongside cosmic web emission so as to provide robust constraints on possible contamination within the stacking signal.

In response to the need for such simulations, we present the first sky model providing both the synchrotron cosmic web alongside a realistically clustered AGN and SFG radio population, the `FIlaments \& GAlactic RadiO' (FIGARO) simulation. We provide this model in the form of ten \SI{4 x 4}{\degree} light cones out to a redshift of $z = 0.8$, and valid for observing frequencies ranging from \SIrange[range-units=single]{100}{1400}{\mega \hertz}. To do this, we have combined cosmic web emission extracted from the magneto-hydrodynamic (MHD) simulations by \citet{Vazza2019} and have populated the light cone with AGN and SFG radio sources using the Tiered Radio Extragalactic Continuum Simulation (T-RECS) codebase produced by \citet{Bonaldi2019}. Uniquely, this latter population are positioned and clustered realistically with respect to the underlying mass density of the cosmological simulation. We expect this simulation to be important in developing observing and detection strategies for detecting the cosmic web with both current as well as upcoming low to mid frequency radio telescopes.

This paper is structured as follows. We begin by discussing the construction of this simulation as well as the verification steps taken during that process. In \autoref{sec:cosmosnaps} we discuss the underlying MHD simulation, the extraction of synchrotron cosmic web emission from the snapshots and the calibration of this emission against the small known population of radio relics. In \autoref{sec:halos}, we discuss the extraction and validation of dark matter halos from the simulation, taking care to verify their number density and clustering properties as this population are a key input for the T-RECS simulation. \autoref{sec:lightcone} details the light cone construction process itself, stacking the cosmic web and dark matter halos out to a redshift depth of $z = 0.8$. Finally, in \autoref{sec:trecs} we discuss how we use the halo light cones to position the AGN and SFG radio population using the T-RECS simulation codebase, before presenting the completed simulation catalogue in \autoref{sec:results}. We finish in \autoref{sec:discussion} with a discussion oriented around the possibility of detection of the cosmic web with the latest generation of radio telescopes.

Throughout this paper, and including the original cosmological simulation, a $\Lambda$CDM cosmological model is assumed, with density parameters $\Omega_{\text{BM}} = 0.0478$ (baryonic matter), $\Omega_{\text{DM}} = 0.2602$ (dark matter), and $\Omega_{\Lambda} = 0.692$, and the Hubble constant $H_0 =$ \SI{67.8}{\kilo \metre \per \second \per \mega \parsec}. All stated observing resolutions refer to the full width, half maximum (FWHM) of a circular Gaussian.

\section{Cosmological simulation \& snapshots}
\label{sec:cosmosnaps}

In beginning to build FIGARO, we must start with an underlying and evolving cosmological simulation. We have used the MHD simulation detailed in \citet{Vazza2019}, which was produced using \textsc{enzo}.\footnote{\url{www.enzo-project.org}} \footnote{Note that the cosmological parameters are incorrectly stated in \citet{Vazza2019}, and are instead those given in this paper.} This simulation encompassed a comoving volume of \SI[parse-numbers=false]{100^3}{\mega \parsec \cubed} with a uniform grid of $2400^3$ cells and $2400^3$ dark matter particles, each with a fixed mass set at \SI{8.62E6}{\solarmass}. This gives a spatial resolution of \SI[parse-numbers=false]{41.6^3}{\kilo \parsec \cubed} (comoving) per cell. The simulation was initialized at $z = 45$ with a simple uniform magnetic field of $B_0 =$ \SI{0.1}{\nano \gauss}, and these fields were evolved in time using the MHD method of \citet{Dedner2002}.

Whilst larger, purely dark matter simulations do exist, a full MHD treatment is significantly more computationally expensive. The simulation used here is the largest of its kind to date, and presents a careful computational trade-off between cellular resolution and total volume. In the latter case, the volume is large enough to allow the simulation of a small population of $\sim$\SI{E14}{\solarmass} clusters and galaxy groups.

An important caveat to note is that the simulation does not include radiative gas cooling, feedback processes due to star formation, or AGN that are important to the evolution of cluster interiors, nor was the spatial resolution sufficiently high to capture turbulent dynamo effects of the dense cluster regions that can significantly magnify magnetic fields. Our regions of interest for this catalogue, however, are the filaments and cluster peripheries and these regions are sufficiently distant from core cluster environs that these effects, at least to first order, are not especially relevant.

\subsection{Radio}

Synchrotron radio emission is produced by electrons at relativistic energies interacting with background magnetic fields. In this simulation we trace synchrotron emission solely as a result of diffusive shock acceleration (DSA) typically associated with accretion shocks, cluster mergers and other large-scale structure formation processes (e.g. \citealp{Ryu2003}).

In DSA, a small fraction of ambient electrons are accelerated to relativistic energies, which then radiate due to their interaction with non-neglible intracluster magnetic fields. We model the resulting radio power based on \citet{Hoeft2007}, henceforth HB07:
\begin{equation}
    P_\nu \propto S \cdot n_d \cdot \xi\left(M,T\right) \cdot \nu^{\alpha} \cdot T_d^{\nicefrac{3}{2}} \cdot \frac{B^{1 - \alpha}}{B^2_{\text{CMB}} + B^2}
\end{equation}
where: $S$ is the shock surface area; $n_d$ is the downstream electron density; $\xi\left(M, T\right)$ is the electron acceleration efficiency which is a function of Mach number and temperature; $\nu$ is the frequency and $\alpha$ is the spectral index of the radio emission; $T_d$ is the downstream electron temperature; $B$ is the magnetic field in each cell; and $B_\text{CMB}$ is the equivalent magnetic field strength of the CMB where $B_\text{CMB} \approx 3.25(1 + z)^2$ \SI{}{\micro \gauss}.

Both $B$ and $\xi\left(M, T\right)$ are poorly understood, especially in the highly diffuse filaments and cluster outskirts, yet are key parameters in calculating the radio power. The magnetic field strength along filaments is primarily the result of adiabatic gas compression, and in these low density regions the field strength is closely related to the magnetic field seeding scenario \citep[e.g.][]{Vazza2015}. The seed strength, however, is unknown and thus predictions of these magnetic fields vary between cosmological simulations in the range of \SIrange[range-phrase=--,range-units=single]{E-4}{0.1}{\nano \gauss}. At best we have upper limits provided by Planck from CMB observations which limit the magnetic field strengths to a few nG on scales of \SI{1}{\mega \parsec} \citep{Planck2016}. In this simulation, the magnetic field was seeded uniformly at \SI{0.1}{\nano \gauss}---an order of magnitude lower than these limits---but there is latitude in the choice of this parameter. Moreover, for values of $B \ll B_\text{CMB}$, the radio power scales $P \propto B^2$, making the simulation particularly sensitive to the seed strength.

The electron acceleration efficiency $\xi\left(M, T\right)$ is a second, crucial parameter that is difficult to model. This parameter estimates the fraction of thermal electrons that are accelerated to relativistic energies as a result of a shock. The model used here is based on HB07 which depends upon the strength of the shock and the thermal temperature of the downstream electrons. However, this model does not exceed $\sim10^{-3}$ for reasonable Mach values, and this value is insufficient to account for many observed radio relics (e.g. \citealp{Botteon2020}). For these events, we now believe `fossil' electron populations---those which have been previously accelerated by, for example, AGN activity or previous DSA shocks---allow for a much higher effective electron acceleration efficiency (e.g. \citealp{Pinzke2013}). As shocks were here calculated in post-processing, this simulation does not have a `memory' of previous shocks nor does it model AGN activity; at each snapshot, the electron population is therefore always assumed to be at thermal equilibrium resulting in underestimated radio emission, especially in dense cluster environments.

As a mitigation, we introduce a modified HB07 model (herein: `HB07+fossil') for the acceleration efficiency. This mitigation builds upon HB07 with a special case for weak shocks in dense environments: for shocks with $\mathcal{M} < 5$ and thermal temperature $T > 10^7$\SI{}{\kelvin}, we arbitrarily set $\xi\left(M, T\right) = 10^{-2}$. The final HB07+fossil model was then a weighted sum of the original HB07 model with this special case, with weightings 0.95 and 0.05 respectively.

\begin{figure*}
    \centering
    \begin{subfigure}{0.8\linewidth}
        \includegraphics[width=\linewidth]{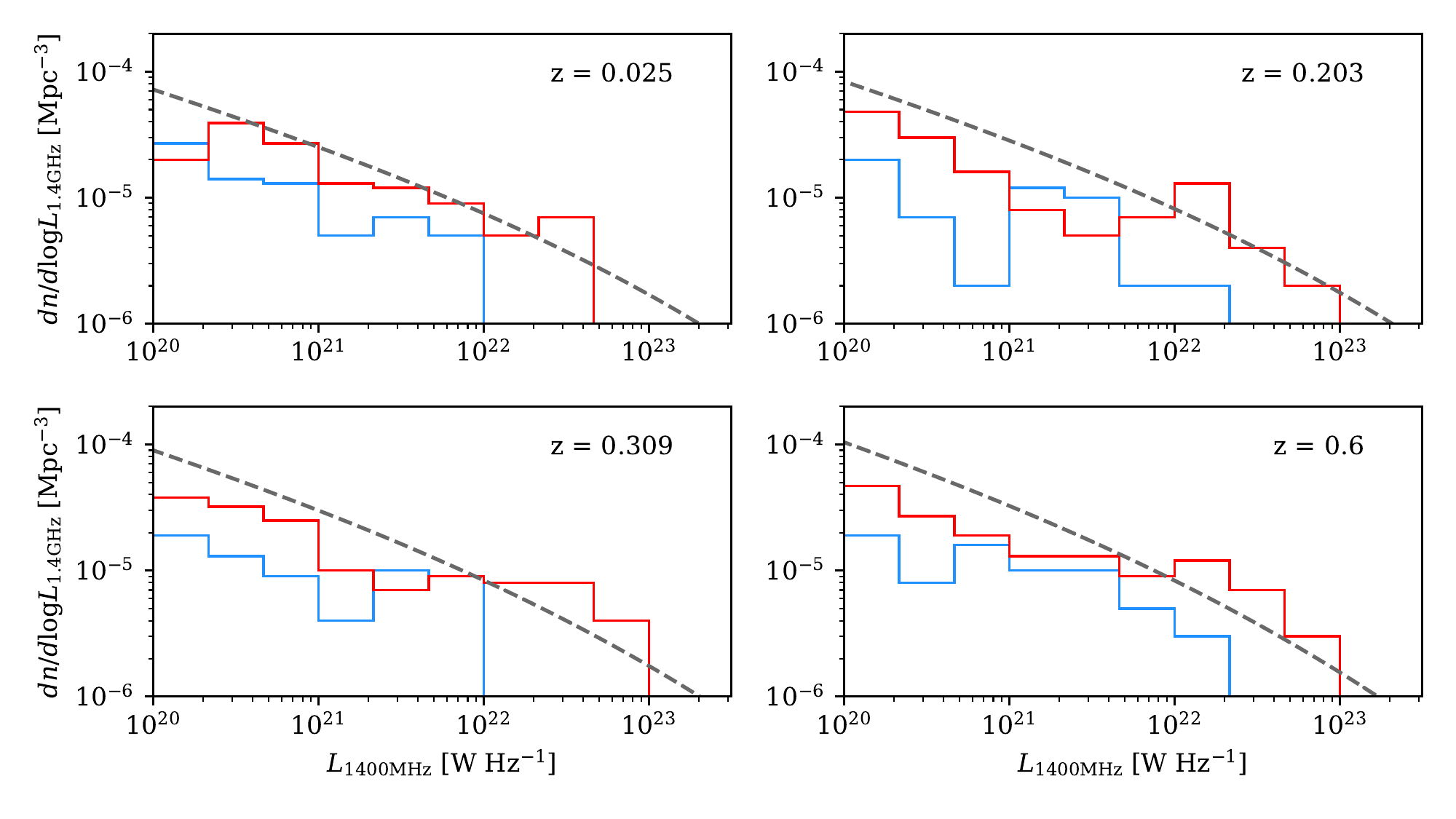}
    \end{subfigure}
    
    \caption{A comparison of the radio relic luminosity function (RRLF) from \citet{Nuza2012} (dashed grey) with the the measured relics in each snapshot volume for HB07 (blue) and HB07 with additional fossil electrons (red). Relic statistics were calculated by summing emission in the annulus around dark matter halos with radii $0.5 \cdot r_{200} < r < 1.5 \cdot r_{200}$.}
    \label{fig:rrlf}
\end{figure*}

These weightings were chosen to best reproduce the radio relic luminosity function (RRLF) derived by \citet{Nuza2012}. Radio relics are a known class of radio source produced by DSA at the periphery of clusters and driven by shocks from large scale structure formation processes. Radio relics should be well modelled by our simulation and provide a means by which to calibrate the radio emission of our models. In \autoref{fig:rrlf} we compare the relic counts for both the HB07 (blue) and HB07+fossil (red) models for each of the snapshot volumes of our simulation\footnote{Relics were measured by summing the luminosity about dark matter halos in the annular cylinder defined with radii $0.5 \cdot r_{200} < r < 1.5 \cdot r_{200}$ and depth $3 \cdot r_{200}$. To avoid double counting in the case of nearby relics, we processed halos in order of most massive to least massive, and excluded regions that had already been counted.} with the \citet{Nuza2012} RRLF in dashed grey.\footnote{The calculation of this RRLF required the convolution of a halo mass function, for which we used \citet{Angulo2012}.} We can observe that the standard HB07 model results in significantly under-powered relics by about one order of magnitude, whilst the modified HB07+fossil weighted model brings the relic counts into good agreement.

\subsection{Snapshots}

\begin{table}
	\centering
	\begin{tabular}{lcc}
		\hline
		\textbf{Snapshot No.} & \textbf{Redshift} & \textbf{Luminosity density (1.4~GHz)} \\
		& z & [erg s$^{-1}$ Hz$^{-1}$ Mpc$^{-3}$]\\
		\hline
		188 & 0.025 & \SI{4.45E24}{} \\
		166 & 0.2 & \SI{6.20E24}{} \\
		156 & 0.3 & \SI{3.58E24}{} \\
		122 & 0.6 & \SI{4.44E24}{} \\
		\hline
	\end{tabular}
	\caption{The simulation snapshots used to construct the light cone and associated halo catalogues.}
	\label{table:snapshots}
\end{table}

We have extracted four `snapshots' of the simulation at redshifts ranging from $z = 0.025$ to $z = 0.6$, as detailed in \autoref{table:snapshots}, and these snapshots have formed the basis for constructing our light cones. This selection of snapshots was driven purely by the data still available. Amongst the observables produced by the simulation, the three of note to this work are: dark matter density cubes; the calculated luminosity of the synchrotron emission at \SI{1400}{\mega \hertz}; and the shock values (i.e. cell Mach numbers).

\section{Halo finding}
\label{sec:halos}

\begin{figure*}
    \centering
    \includegraphics[width=\linewidth]{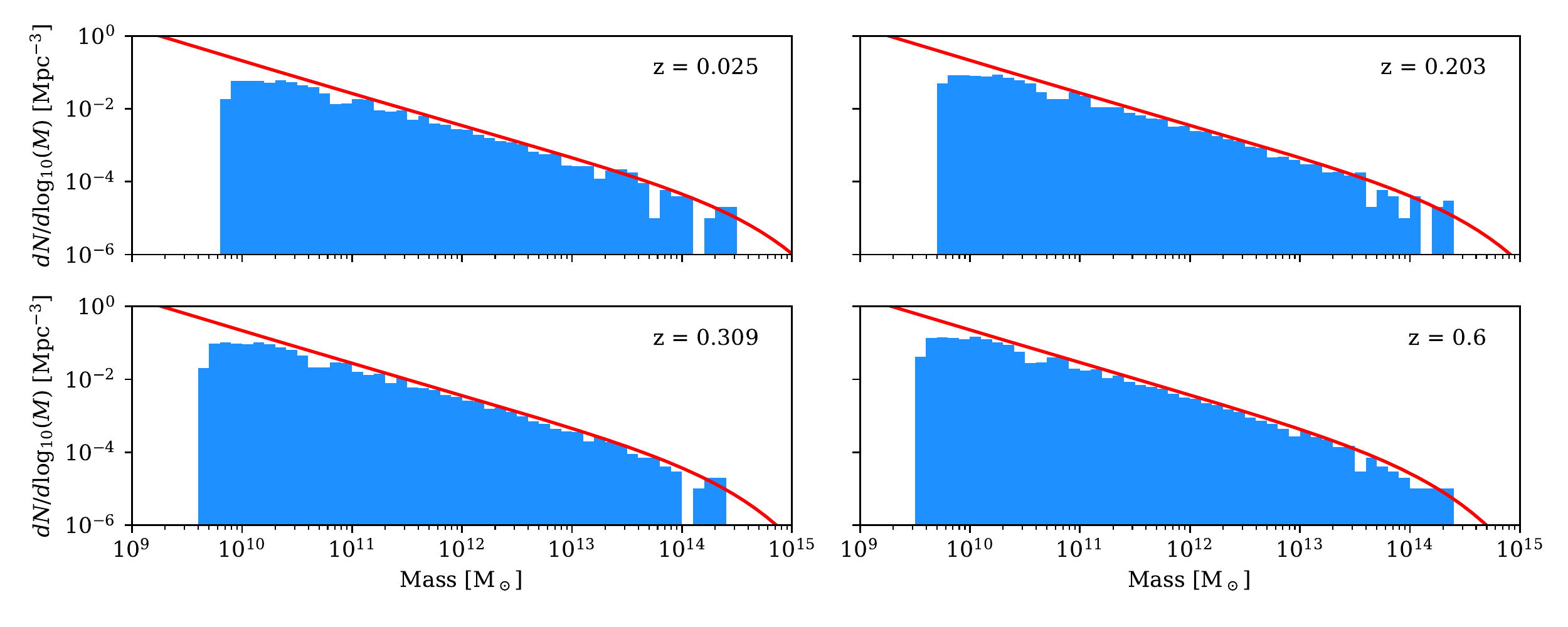}
    \caption{Halo counts for each snapshot, compared with the halo mass function by \citet{Angulo2012} (produced using HMFCalc by \citealp{Murray2013}) at the respective redshifts. The halos are binned by mass into intervals of 0.1 dex.}
    \label{fig:halomassfunction}
\end{figure*}

\begin{figure*}
    \centering
    \begin{subfigure}{0.49\linewidth}
        \includegraphics[width=\linewidth]{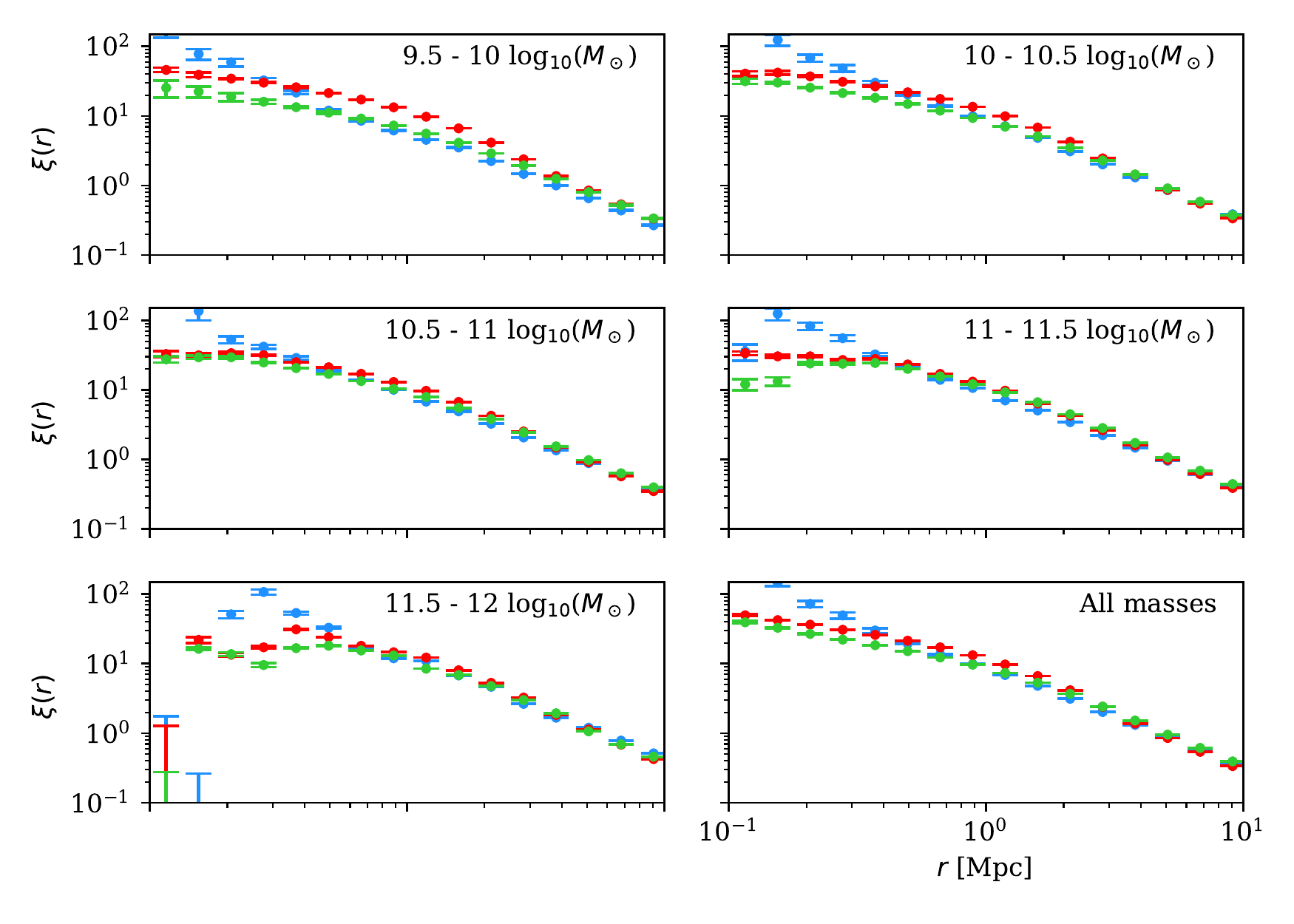}
        \caption{$z \approx 0.025$}
        \label{fig:halo2point:a}
    \end{subfigure}
    \begin{subfigure}{0.49\linewidth}
        \includegraphics[width=\linewidth]{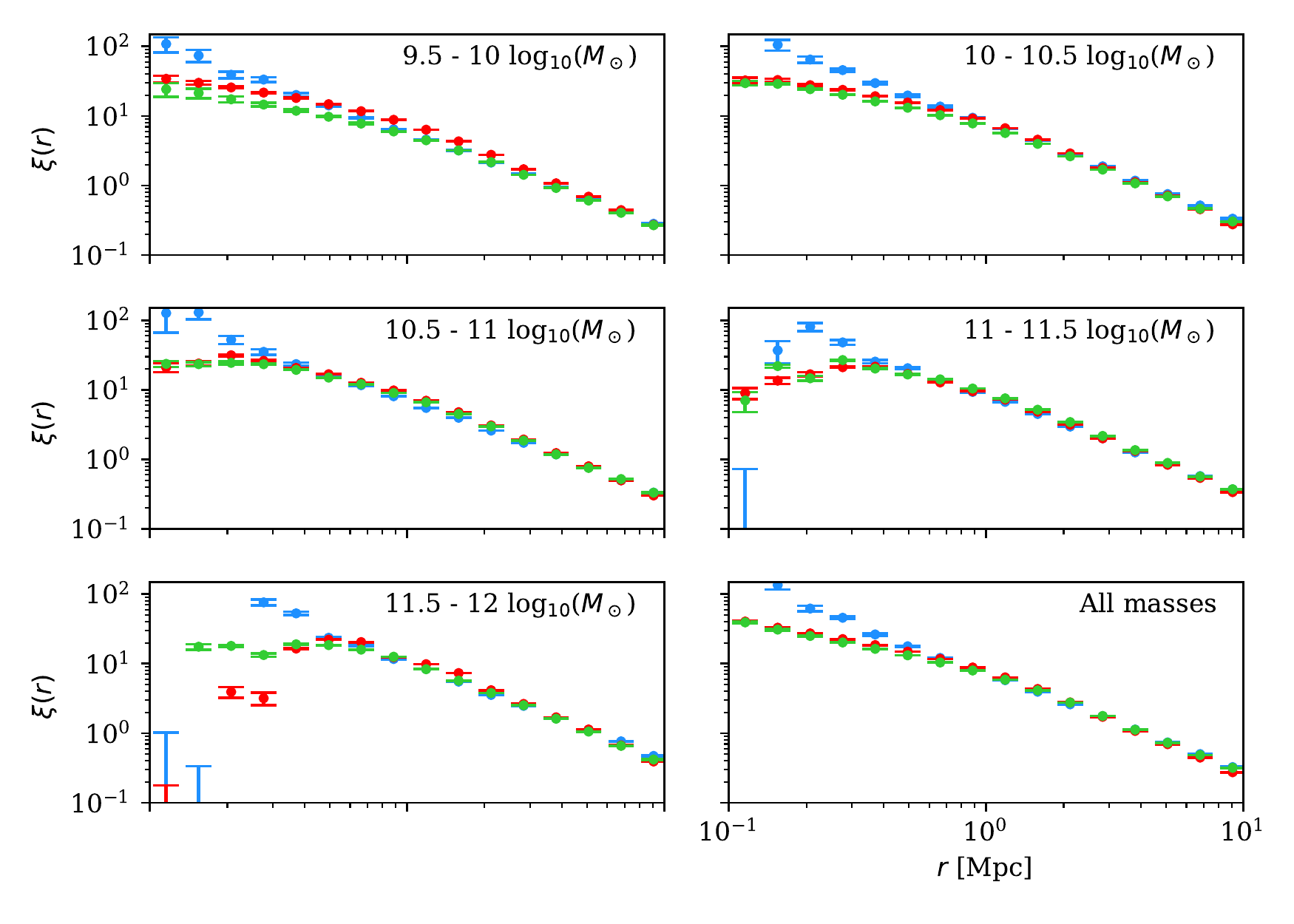}
        \caption{$z \approx 0.309$}
        \label{fig:halo2point:b}
    \end{subfigure}
    \caption{The two-point correlation of halos for the FIGARO (blue), Millennium II (MII; red) and Planck Millennium (PM; green) simulations. The results are shown at redshifts $z \approx 0.025$ and $z \approx 0.309$ and are binned by mass in 0.5 $\text{log}_{10}$(\SI{}{\solarmass}) increments. Additionally, an all mass result is also shown. In all, the halo clustering properties of FIGARO are consistent with both MII and PM simulations with the exception of scales under $\sim$\SI{0.3}{\mega \parsec}.}
    \label{fig:halo2point}
\end{figure*}

Alongside the radio emission, the mass distribution, and in particular the dark matter halo positions, are a key ingredient in building the FIGARO simulation. This arises because T-RECS relies on dark matter halos to position AGN and SFG and thus accurately mimic the clustering properties of these sources with respect to the cosmic web emission. Thus for each snapshot, we needed to identify dark matter halos in post-processing using each snapshot's associated dark-matter cube. The processed outputs of the simulation have been smoothed using a Cloud In Cell (CIC) kernel, prohibiting traditional `friend of friend' algorithms. Instead, to identify dark matter halos we used the following simple algorithm: a sphere was progressively grown around the most massive voxel until the mean density of the enclosed volume reduced beneath a threshold density $\bar{\rho}$, in this case $\bar{\rho} = 200 \rho_c$, where $\rho_c$ is the critical density of the Universe at that respective redshift. From this we could then interpolate the virial radius $r_{200}$ and mass $M_{200}$ of the halo. The mass of enclosed voxels was set to zero and the process was repeated on the next most massive voxel in the snapshot, and so on, until exhausting the volume of resolved halos.

To provide enough potential source positions, T-RECS requires that we find halos down to $\sim$\SI{E9.5}{\solarmass}. Given the relatively coarse spatial and mass resolution of our simulation, detecting these low mass halos is challenging. To achieve this we set the minimum allowable halo size as having a radius of 1 voxel, that is, incorporating just 7 voxels in total. At this limit, however, we are especially wary of the introduction of errors or bias due to discretisation effects, and we have therefore performed a number of sanity checks upon our halo catalogue.

In \autoref{fig:halomassfunction} we show the halo counts, binned by mass, for each respective snapshot. We compare these counts with the halo mass function by \citet{Angulo2012} (produced using HMFCalc by \citealp{Murray2013}) and we observe good agreement down to about \SI{E10.5}{\solarmass}, below which our catalogue becomes increasingly incomplete. The lowest mass halos range from \SI{E9.87}{\solarmass} at $z = 0.025$ to \SI{E9.57}{\solarmass} at $z = 0.6$, where this lower limit steadily decreases due to the decreasing cosmological scale factor and corresponding increased resolution of the simulation at earlier epochs.

As a second sanity check, in \autoref{fig:halo2point} we compare the clustering properties of the FIGARO halos with those from the Millennium II (MII) and Planck Millennium (PM) simulations by calculating the two-point correlation functions of each respective halo population (for details, see \autoref{appendix:twopoint}). Both the MII and PM simulations are of sufficiently high resolution so as to be able to properly resolve low mass halos on the order of \SI{E9.5}{\solarmass} and therefore provide a good benchmark for comparison. What we observe is good agreement between the simulations across all mass ranges for scales greater than \SI{0.3}{\mega \parsec}. For scales less than this, both MII and PM turnover slightly whereas FIGARO continues its power law behaviour. This discrepancy exists for all mass bins and so is not directly related to the discretisation of the simulation, although its cause is unknown.

These sanity checks satisfy to us that our halo catalogue is sufficiently robust, despite the coarse resolution of our simulation.

\section{Light cone construction}
\label{sec:lightcone}

With both the synchrotron cosmic web and the dark matter halos catalogued for each of our underlying snapshot volumes, we can now proceed to tile these volumes so as to construct the FIGARO light cones. This cone was constructed spanning a \SI{4 x 4}{\degree} angular extent and with a depth extending to a redshift of $z = 0.8$. The simulation volume, with sides of length \SI{100}{\mega \parsec} (comoving) only extends out to $z = 0.023$, and the field of view exceeds a comoving transverse distance of \SI{100}{\mega \parsec} at $z = 0.35$. This necessitates replicating the simulation volume along the length of the light cone a total of 34 times, as well as laterally tiling the volume beyond $z = 0.35$. We discuss here the construction of this light cone.

The web and halo light cones were constructed identically. In both cases, we began with a catalogue of voxel positions and properties. For the web, these properties were the voxel luminosity and shock value; for the halos these were the halo mass and radius. Each cone was constructed by appending the snapshot catalogue in 100 Mpc increments along the length of the light cone. To mitigate repeating structures along the line of sight, we introduced a random transverse offset on each iteration as well as sequentially rotating the volume through each of its three axes. Then, for each entry in the snapshot catalogue, we calculated its redshift, latitude and longitude and, if these latter values fit within the \SI{4 x 4}{\degree} field of view, they would be appended to the final catalogue. Note that throughout, we assume the field of view is small enough that we can safely assume a flat screen projection.

As noted, at redshifts of about $z = 0.35$ it becomes necessary to tile the simulation volume so as to fully cover the field of view. Since the simulation volume wraps at its edges, this tiling does not cause discontinuities or edges in the final light cone. The volume was tiled in a $2\times2$ arrangement from $z = 0.35$.

The random transverse offset allows us to additionally produce multiple light cones or `realisations' which, especially at low redshifts, can be significantly different to each other. In one realisation, for example, the low redshift cone may be largely empty whilst in another, just by chance, there may be a massive galaxy cluster within the field of view. This reflects the kind of cosmic variance we should expect. We have produced 10 realisations in total.

In the case of the halos, their properties were simply appended to a catalogue. In the case of the web emission, however, some additional processing was required. For each web voxel, we calculated its flux $S_{\nu}$ from the simulation luminosity values $L_{\nu}$ using the standard radio luminosity function with $k$-correction:
\begin{equation}
    S_{\nu}\left(z\right) = L_{\nu} \frac{(1 + z)^{\alpha}}{4 \pi D_L^2}
\end{equation}
where $D_L$ is the luminosity distance at the redshift $z$ of the voxel and $\alpha$ is the voxel's spectral index. In turn, the spectral index $\alpha$ was calculated from the voxel's associated Mach number $\mathcal{M}$ using the relation:
\begin{equation}
    \alpha = \frac{\mathcal{M}^2}{1 - \mathcal{M}^2}
\end{equation}
where the power law relation uses the positive sign convention $S \propto \nu^\alpha$. This latter relation is also used to scale the luminosity values from the 1400 MHz output of the simulation to the observing frequency. Finally, this flux value was appended to a map spanning the field of view with \SI{3 x 3}{\arcsecond} resolution.

The relatively small volume of the simulation raises issues as we go deeper in redshift space. Whilst the simulation volume is large enough to give a good representative sample of the Universe for small and medium mass clusters, it is however just large enough to simulate a single massive galaxy cluster, on the order of \SI{E15}{\solarmass}. As we show in \autoref{tab:fractionalvolume}, at low redshifts we sample only a very small fraction of the total simulation volume; since this sample will primarily be low or medium mass clusters it is a reasonable representation of the Universe as whole. Moreover, at these low redshifts the differences between realizations should give an accurate picture of the effect of cosmic variance. However, as we go deeper in redshift and as the fractional volume increases towards unity, the different realizations will increasingly sample from the same parts of the simulation, thus becoming self-similar. Moreover, the particular statistics associated with the small population of high mass clusters in our volume, which also host the most luminous cosmic web emission, will begin to dominate statistics derived from these redshift slices. Finally, at redshifts of around $z = 0.35$, where tiling becomes necessary, the most massive galaxy cluster in the volume will be replicated in every single \SI{100}{\mega \parsec} increment. This single cluster and its unique evolution will dominate most statistics derived from these deeper redshift slices. This is a limitation we will need to consider when we later discuss the statistics of cosmic web emission.

\begin{table}
    \centering
    \begin{tabular}{p{0.1\linewidth}<{\centering}p{0.1\linewidth}<{\centering}p{0.25\linewidth}<{\centering}p{0.25\linewidth}<{\centering}} \toprule
        \multicolumn{2}{c}{Redshift slice} & Comoving volume &  Fractional volume  \\
        $z_\text{min}$ & $z_\text{max}$ & [Mpc$^3$] &  \\ \midrule
        0 & 0.05 & \SI{1.7E4}{} & 0.02 \\
        0.05 & 0.1 & \SI{1.1E5}{} & 0.11 \\
        0.1 & 0.15 & \SI{3.0E5}{} & 0.30 \\
        0.15 & 0.2 & \SI{5.5E5}{} & 0.55 \\
        0.2 & 0.25 & \SI{8.6E5}{} & 0.85 \\
        0.25 & 0.3 & \SI{1.2E6}{} & 1.21 \\
        0.3 & 0.35 & \SI{1.6E6}{} & 1.60 \\ \bottomrule
    \end{tabular}
    \caption{The comoving volume enclosed by regular $\Delta z = 0.05$ redshift slices and the \SI{4 x 4}{\degree} field of view. For higher redshifts ($z \gtrapprox 0.15$), the volume as a fraction of the total simulation volume is sufficiently large enough that different realisations will be increasingly similar. For redshift slices greater than approximately $z = 0.25$, it becomes necessary to duplicate the simulation volume more than once.} 
    \label{tab:fractionalvolume}
\end{table}

\section{T-RECS}
\label{sec:trecs}

T-RECS \citep{Bonaldi2019} is a simulation of the continuum radio sky from \SI{150}{\mega \hertz} to \SI{20}{\giga \hertz} that models radio emission from AGN and SFG using a number of empirically derived relations, and forms the final ingredient required for FIGARO. The simulation provides a thorough range of properties for each radio source, ranging from general properties of intrinsic luminosity, redshift, and physical size to more specific properties such as AGN classification, radio jet angle, and SFG ellipticity. Each of the two general populations were further broken down into subpopulations. The AGNs consisted of: steep-spectrum sources (SS-AGNS), flat-spectrum radio quasars (FSRQ), and BL Lac. The SFGs were also further subdivided into three subpopulations: late-type, spheroidal, and lensed spheroidal galaxies.

The T-RECS simulation was constructed with careful attention paid to the spatial clustering of radio sources. This clustering was implemented by associating each AGN and SFG radio source with a dark matter halo extracted from a \SI{5 x 5}{\degree} light cone that \citet{Bonaldi2019} had constructed, originally derived from the PM simulation. \citet{Bonaldi2019} describe in detail the way in which each radio population was associated with dark matter halos of a particular mass range, which we briefly summarize here. In the first case, the AGN subpopulations were associated with dark matter halos by first relating a stellar mass $M_*$ to each halo mass $M_h$ (i.e. $M_* = F(M_h)$), and then in turn calculating the fraction of galaxies hosting an AGN (radio loud or radio quiet) as a function of that host galaxy mass. The SFGs, on the other hand, used an abundance matching process to relate the known distribution of halo masses and the known distribution of SFG luminosities. This allowed for associating the most luminous SFGs with dark matter halos, whilst for SFGs where the luminosity implied a dark matter halo mass smaller than allowed for by the resolution of the PM simulation, a random distribution on the sky was instead assumed.

Crucially for our purposes, \citet{Bonaldi2019} also made the simulation code publicly available. The has allowed us to run the simulation ourselves, and in the process input our own dark matter halo catalogue for each of the respective realisation light cones. Thus the output extragalactic catalogue positions AGN and SFG sources with respect to the underlying matter density of our lightcones, and in this way we accurately model how the cosmic web and the embedded radio population cluster with respect to each other.

We made use of the T-RECS codebase largely without modification, with the exception of a bug fix that corrected a random number generation routine that incorrectly produced strongly biased results. We provided the halo catalogues of each light cone realisation in the format T-RECS expected and ran the simulation for a \SI{4 x 4}{\degree} field of view, with a lower flux threshold of $S_\text{1400 MHz} =$ \SI{0.1}{\micro \jansky}, and for a redshift range that encompassed the full cosmic web light cone. The choice of lower flux threshold was selected so as to fully simulate classical confusion noise in the frequency range \SIrange[range-phrase=--,range-units=single]{150}{1400}{\mega \hertz} assuming a highest resolution of \SI{8}{\arcsecond} (see Appendix \autoref{appendix:confusion}).

Finally, we also note the ease with which future improvements to T-RECS modelling of AGN and SFG populations can readily be regenerated into this simulation, especially as deeper, large-field surveys provide more accurate statistics of these faint sources \citep[e.g.][]{Prandoni2021}.

\subsection{Clustering of the radio population}

\begin{figure*}
    \centering
    \includegraphics[width=\linewidth]{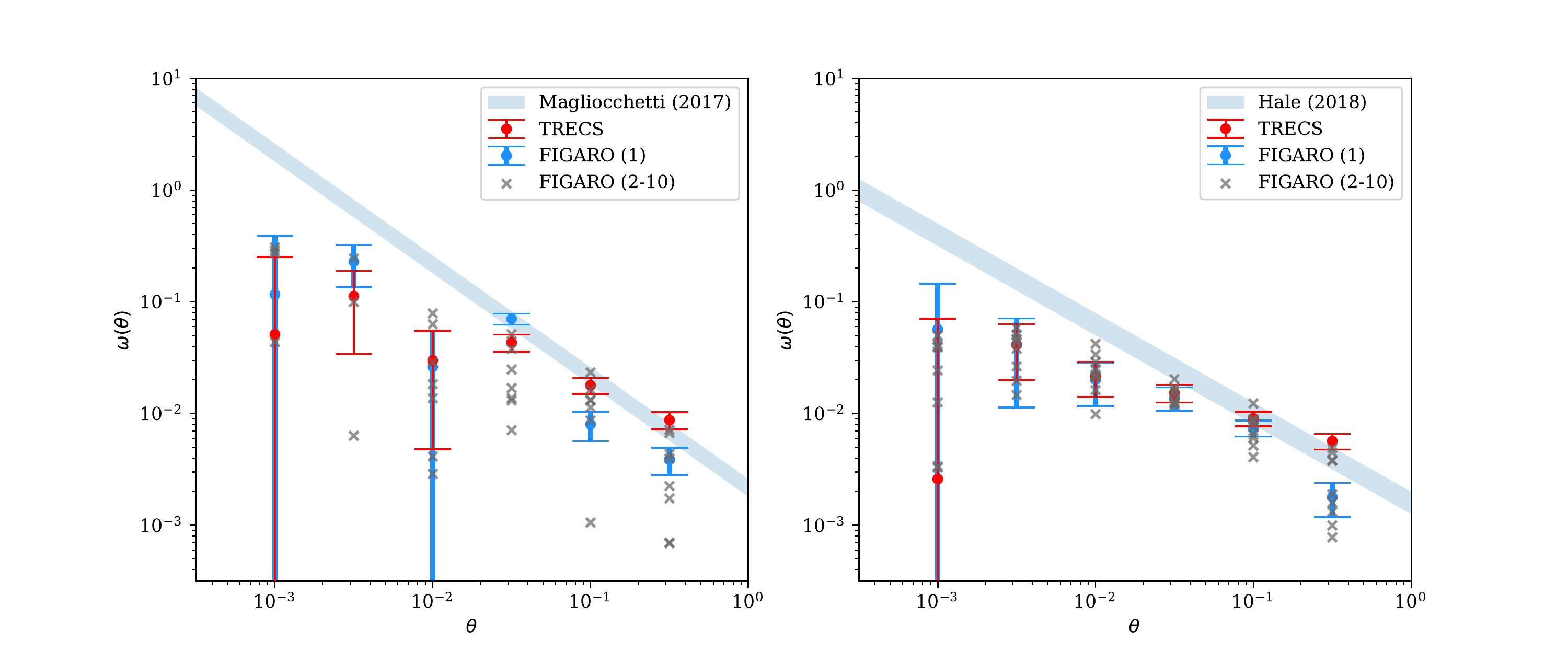}
    \caption{The angular two-point correlation of all radio sources (AGN, SFG and all subtypes) for FIGARO (realization 1; blue), the original TRECS (red), and other FIGARO realizations (gray crosses). \textit{Left:} Comparison to \citet{Magliocchetti2017} using their minimum flux threshold of \SI{0.15}{\milli \jansky} at \SI{1400}{\mega \hertz}. \textit{Right:} Comparison to \citet{Hale2018} using their minimum flux threshold of \SI{12.65}{\micro \jansky} at \SI{3}{\giga \hertz}}. 
    \label{fig:angular2point}
\end{figure*}

To validate the spatial distribution of the resulting radio catalogue, we turn to the the angular two-point correlation function. This function $\omega\left(\theta\right)$ describes the apparent two-dimensional clustering of radio sources based on their angular separation, without reference to their redshift. Here we compare the results of two recent empirical measurements of this value, by \citet{Magliocchetti2017} and \citet{Hale2018}, against that of our simulation.

\citet{Magliocchetti2017} reported a measurement of the angular two-point correlation function based on a catalogue of sources derived from observations at \SI{1.4}{\giga \hertz} of a 2~deg$^2$ region of the COSMOS field. By setting a lower flux threshold of \SI{0.15}{\milli \jansky}, above which value the catalogue was considered complete, and assuming $\omega\left(\theta\right)$ to be a power law of the form $\omega\left(\theta\right) = A \theta^{1 - \gamma}$ where $\gamma$ was set to 2, they were thereby able to derive a value for the constant of proportionality as $A =$~\SI{2.2(4)E-3}{}. \citet{Hale2018} similarly used observations of a 2~deg$^2$ region of the COSMOS field, but at the higher frequency range of \SIrange[range-phrase=--,range-units=single]{2}{4}{\giga \hertz} and using a much lower flux threshold \SI{12.65}{\micro \jansky} (5.5 times the median image noise). They assumed the same power law form but instead fixed $\gamma = 1.8$ and thereby derived a value for the constant of proportionality as $\log_{10} A =$ \SI{-2.8(1)}{}.

In \autoref{fig:angular2point} we compare these measurements against those of our simulation. We estimate the angular version of the two-point correlation function using the equation described in \autoref{appendix:twopoint} with the exception that the Euclidean metric $r$ is replaced with the apparent angular separation $\theta$ between sources. All error estimates are calculated using bootstrap sampling. The function was calculated for angular separations $10^{-3.25} < z < 10^{-0.25}$ in equal logarithmically spaced bins of width $\log_{10} \Delta \theta = 0.5$.

In the left panel, we compare the measurement of \citet{Magliocchetti2017} (shaded blue region) with those of realization 1 from FIGARO (blue, error bars indicate 3 standard deviations). We also calculate $\omega\left(\theta\right)$ for the other realizations (grey crosses), which indicate the variation in this function purely as a result of cosmological variance; however, for clarity we have not included their associated error bars. For reference, we also plot $\omega(\theta)$ of T-RECS using its original catalogue release.\footnote{We note that these two point statistics have been recalculated by ourselves and are notably lower than those published by \citet{Bonaldi2019}.} The simulated catalogues have had a flux threshold of $S_\text{1.4 GHz} >$ \SI{0.15}{\milli \jansky} applied. On the right, we compare the measurement of \citet{Hale2018} with the same set of simulations except subject to a flux threshold of $S_\text{3 GHz} >$ \SI{12.65}{\micro \jansky}.

In both cases, FIGARO is reasonably consistent with these empirical measurements for $\theta \gtrapprox 10^{-1.75}$, especially when we take into account the spread of values measured from different realizations, as well as with the original T-RECS. FIGARO (and the original T-RECS), however, appears to be under-clustered on angular scales smaller than this. One possible explanation for this is a result of our coarse simulation volume. The resolution of our simulation (and therefore the halo catalogue) was \SI{41.7}{\kilo \parsec}, which at a redshift of $z = 0.1$ corresponds to a minimum angular resolution of \SI{19.9}{\arcsecond} and even at a redshift of $z = 0.3$ the minimum angular resolution of \SI{7.0}{\arcsecond} is still greater than the smallest bin for which we have calculated $\omega(\theta)$. As a result, for a given redshift slice, we should expect $\omega(\theta)$ will decline for values of $\theta$ smaller than the respective minimum angular scale, and this will introduce a strong under-clustering bias to our results on these small angular scales. Nonetheless, we consider our results to be at least as good as the original T-RECS catalogue and sufficient for our present purposes.

\section{Results}
\label{sec:results}

\begin{figure*}
    \centering
    \includegraphics[width=0.8\linewidth,clip,trim={1.8cm 3.5cm 0 4cm}]{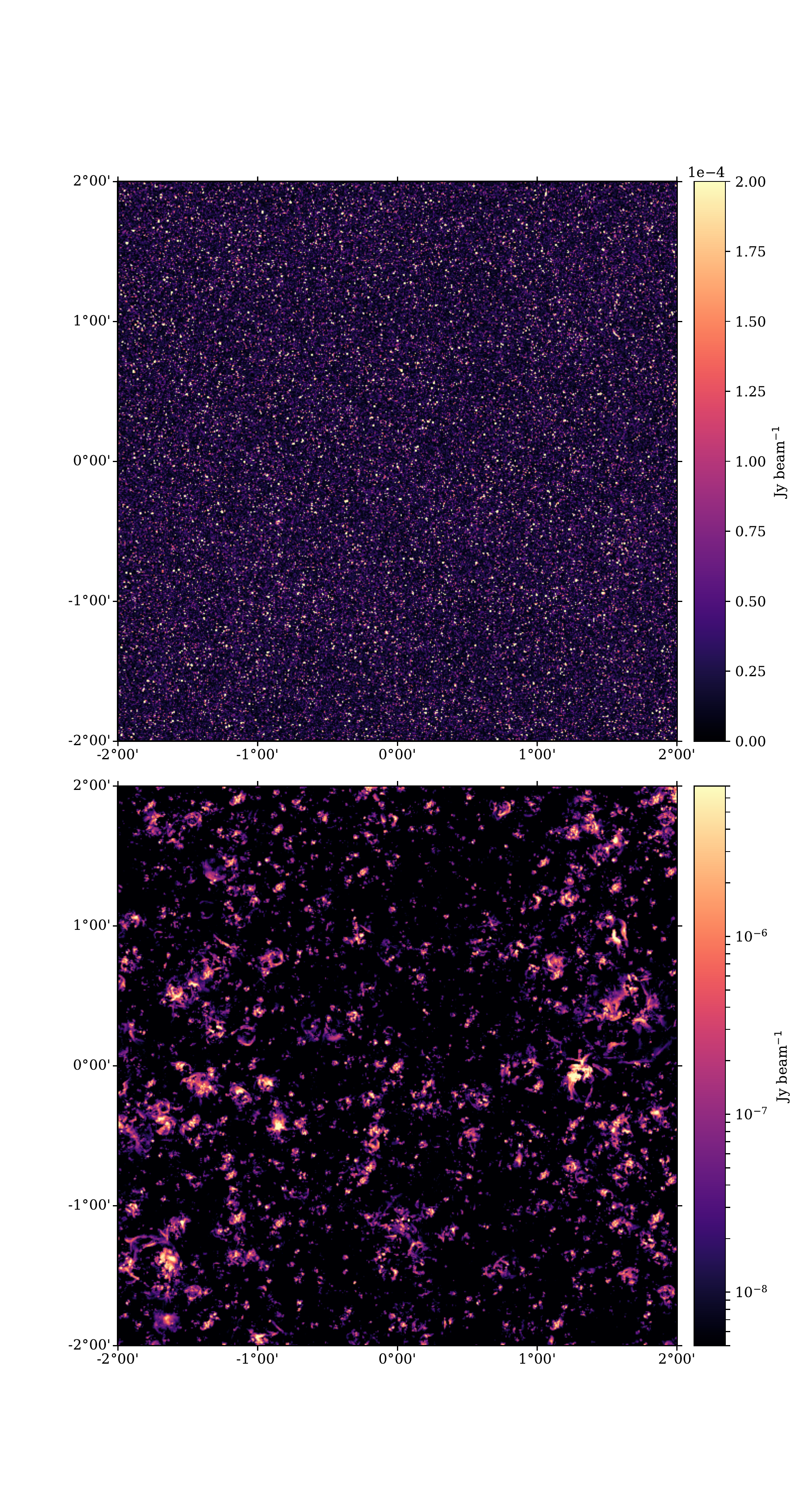}
    \caption{A full field (\SI{4 x 4}{\degree}) image of realisation 5 at \SI{900}{\mega \hertz} encompassing redshift range $0 < z < 0.8$ for the cosmic web and $0 < z < 8$ for T-RECS sources, and convolved to a beam resolution of \SI{20}{\arcsecond}. \textit{Top:} The combined simulation with T-RECS extragalactic sources and faint, background cosmic web emission. The color scale ranges from 0 to saturate at \SI{200}{\micro \jansky}. \textit{Bottom:} Cosmic web emission only, using a logarithmic color scale.}
    \label{fig:allsky}
\end{figure*}

Combining the final cosmic web light cones and T-RECS catalogues, we produce ten \SI{4x4}{\degree} realisations of the radio sky encompassing the redshift range $0 < z < 0.8$. We additionally provide a background catalogue consisting only of T-RECS sources in the redshift range $0.8 < z < 8$; these sources were not clustered but were instead positioned randomly across the field of view using a uniform distribution, which was uniquely generated for each realization. \autoref{fig:allsky} is illustrative of the final cosmic web catalogue, where in the top image we present realisation 5 encompassing the full simulated redshift range of cosmic web emission ($0 < z < 0.8$) and T-RECS sources ($0 < z < 8$) at \SI{900}{\mega \hertz}, and convolved to a beam resolution of \SI{20}{\arcsecond}. The density fluctuations of the extragalactic radio populations are subtly visible, and using this color scale we can also identify a small handful of peaks in the cosmic web emission. In the bottom panel we present the cosmic web emission only, presented using a logarithmic color scale. 

Accompanying this paper, we also make available a full data release of the simulation for each realisation, which includes catalogues describing their respective halos, cosmic web maps, and AGN and SFG populations along the length of the redshift cone. For each realization, we provide the full data in Hierarchical Data Format 5 (HDF5),  with five named datasets. These dataset names and their corresponding schemas are: \texttt{halos} (\autoref{tab:halos}), \texttt{web} (\autoref{tab:web}), \texttt{sfg} (\autoref{tab:sfgs}), \texttt{agn} (\autoref{tab:agns}).

The cosmic web radio emission itself is provided as a 4-dimensional array, with the first axis spanning integrated redshift slices in 0.05 increments, the next  two axes spanning the full \SI{4 x 4}{\degree} field of view, and the final two a tuple containing the flux value and spectral index. The field of view is approximated as a flat screen, with each pixel coordinate offset by \SI{3}{\arcsecond}; with this approximation each pixel occupies approximately \SI{3 x 3}{\arcsecond} with a maximum error at the edge of 0.06\%.

\begin{table*}
    \centering
    \begin{tabular}{p{0.1\textwidth}p{0.2\textwidth}p{0.1\textwidth}p{0.5\textwidth}} \toprule
        Column & Name & Units & Description \\ \midrule
        1 & x\_coord & degree & The x angular coordinate of the halo \\
        2 & y\_coord & degree & The y angular coordinate of the halo \\
        3 & redshift & & The redshift of the halo \\
        4 & R200 & Mpc & The R200 virial (spherical) radius of the halo \\
        5 & M200 & $\log_{10} M_{\odot}$ & The total mass of the volume enclosed the by the R200 virial radius \\ \bottomrule
    \end{tabular}
    \caption{Halo dataset schema which describes the properties of dark matter halos along the length of a realization's redshift cone.}
    \label{tab:halos}
    
    \begin{tabular}{p{0.1\textwidth}p{0.1\textwidth}p{0.1\textwidth}p{0.6\textwidth}} \toprule
        Axis & Length & Name & Description \\ \midrule
        0 & 7 & i & For $i = 0$, this array contains the full sum of cosmic web emission out to redshift $z < 0.8$. The indexes $i \in \{1, 2, ..7\}$ contain the cosmic web emission in redshift slices of depth $\Delta z = 0.05$ spanning $0 < z \leq 0.3$. \\
        1 & 4800 & j & The x angular coordinate of the pixel, where x\_coord = $(j + 0.5) / 1200 - 2$ degrees \\
        2 & 4800 & k & The y angular coordinate of the pixel, where y\_coord = $(k + 0.5) / 1200 - 2$ degrees \\
        3 & 2 & l & The tuple $(S_{\text{900 MHz}}, \alpha)$ where $S(\nu) = S_\text{900 MHz} (\nicefrac{\nu}{\text{900 MHz}})^\alpha$ \\ \bottomrule
    \end{tabular}
    \caption{Cosmic web array with 4 dimensions. The values for x\_coord and y\_coord denote the centre of the pixel, with each pixel having a value of \SI{}{\jansky} and occupying an area $\sim$\SI{3 x 3}{\arcsecond}. The formulas above assume zero-indexing (i.e. $i \in [0, 1, ...])$.}
    \label{tab:web}
    
    \begin{tabular}{p{0.05\textwidth}p{0.2\textwidth}p{0.15\textwidth}p{0.5\textwidth}} \toprule
        Column & Name & Units & Description \\ \midrule
        1 & logSFR & $\log_{10}(M_\odot) / \text{year}$ & Star formation rate \\
        2:7 & I\_{freq} & Jy & Total flux density of the source for frequencies $\nu \in$ (100, 300, 600, 900, 1400, 3000) MHz \\
        8 & x\_coord & degree & The x angular coordinate of the SFG \\
        9 & y\_coord & degree & The y angular coordinate of the SFG \\
        10 & redshift & & The redshift of the SFG \\
        11 & size & arcsecond & Projected apparent size of the disc \\
        12 & e1 & & First ellipticity component \\
        13 & e2 & & Second ellipticity component \\
        14 & PopFlag & & Population flag: 1 $\Rightarrow$ late-type, 2 $\Rightarrow$ spheroidal, 3 $\Rightarrow$ lensed spheroidal \\ \bottomrule
    \end{tabular}
    \caption{SFG dataset schema which describes the properties of SFG radio sources along the length of a realization's redshift cone.}
    \label{tab:sfgs}
    
    \begin{tabular}{p{0.05\textwidth}p{0.25\textwidth}p{0.15\textwidth}p{0.45\textwidth}} \toprule
        Column & Name & Units & Description \\ \midrule
        1 & Lum1400 & $\log_{10}(\text{erg} \text{ s}^{-1} \text{ Hz}^{-1})$ & Luminosity at \SI{1400}{\mega \hertz} \\
        2:7 & I\_{freq} & Jy & Total flux density of the source for frequencies $\nu \in$ (100, 300, 600, 900, 1400, 3000) MHz \\
        8 & x\_coord & degree & The x angular coordinate of the AGN \\
        9 & y\_coord & degree & The y angular coordinate of the AGN \\
        10 & redshift & & The redshift of the AGN \\
        11 & length & Kpc & The physical length of the core plus jet emission \\
        12 & angle & degree & Viewing angle between the jet and line-of-sight \\
        13 & Rs & & Ratio between the distance between the spots and total size of the jets. \\
        14 & PopFlag & & Population flag: 4 $\Rightarrow$ FSRQ, 5 $\Rightarrow$ BL Lac, 6 $\Rightarrow$ SS-AGNs \\ \bottomrule
    \end{tabular}
    \caption{AGN dataset schema which describes the properties of AGN radio sources along the length of a realization's redshift cone.}
    \label{tab:agns}
\end{table*}

\subsection{Morphology \& the ubiquity of `relics'}

\begin{figure}
    \centering
    \includegraphics[width=\linewidth,clip,trim={0 0.5cm 0 0.2cm}]{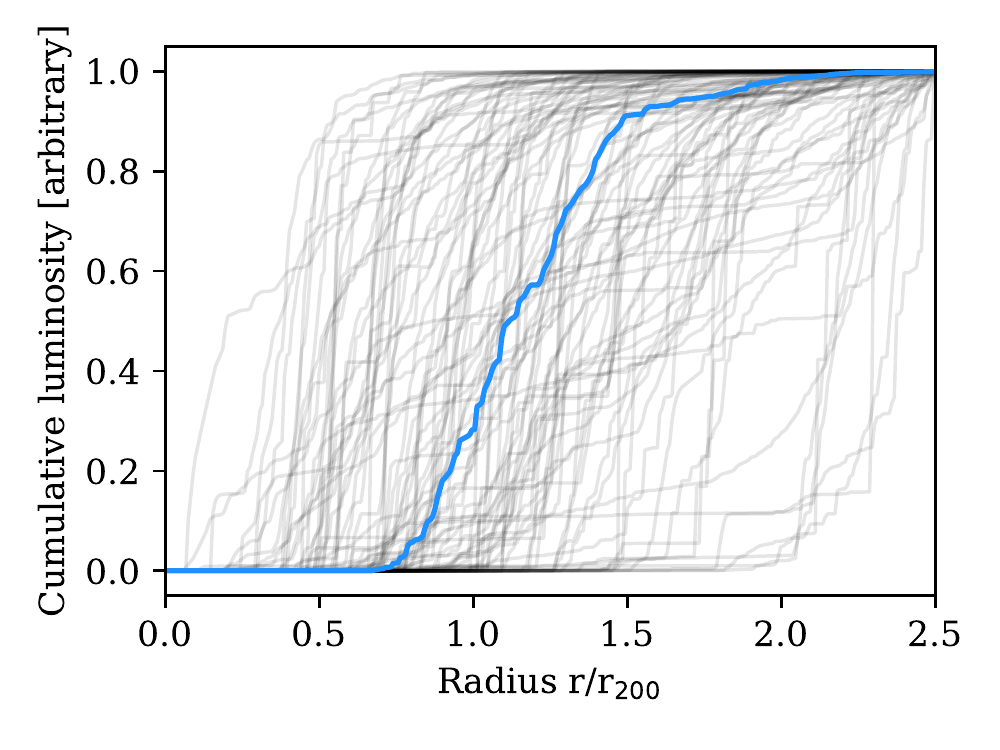}
    \caption{The cumulative radio luminosity contained within the spherical volume surrounding each of the 100 most massive dark matter halos as a function of radius, for the snapshot volume at $z = 0.025$. The luminosity has been normalised to the value at $r = 2.5 \cdot r_{200}$ for each curve. The median value across all halos is indicated in blue.}
    \label{fig:fluxradius}
\end{figure}

\begin{figure}
    \centering
    \includegraphics[width=\linewidth,clip,trim={0 0.5cm 0 0.2cm}]{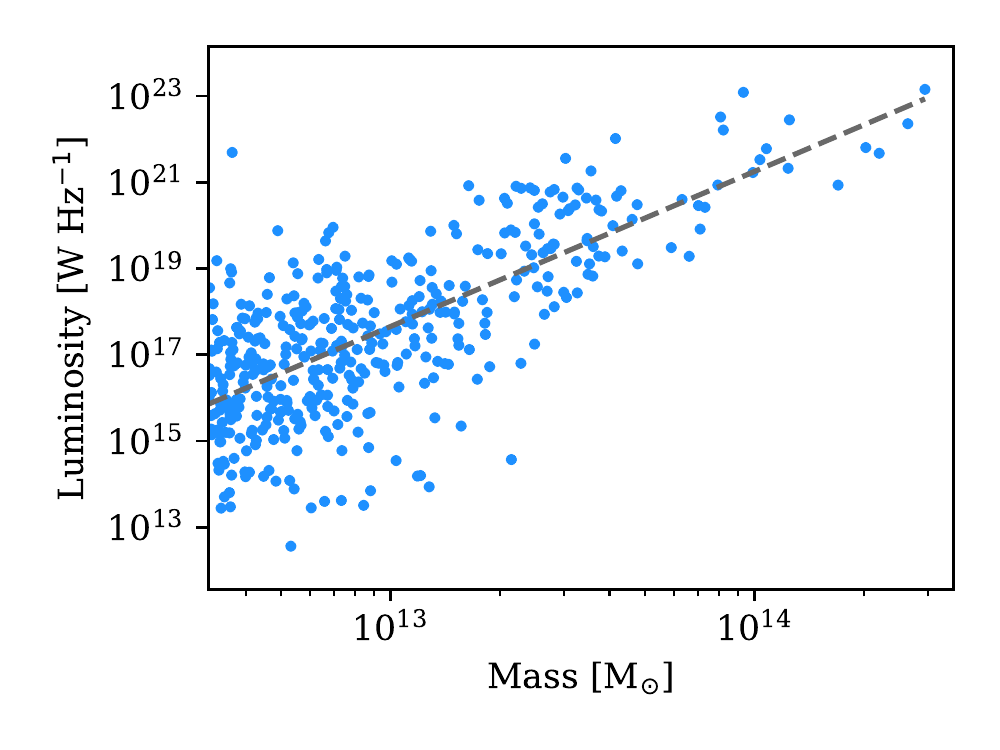}
    \caption{Cosmic web power across the snapshot volume $z = 0.025$ within the dark matter halo spheres $r < 1.5 \cdot r_{200}$ as a function of dark matter halo mass. There is a power law trend indicated by dashed line ($L \propto M^{3.6}$), but significant scattering occurs primarily as a result of the interaction and merger histories of specific clusters.}
    \label{fig:massvspower}
\end{figure}

In \autoref{fig:allsky} we showed the full \SI{4 x 4}{\degree} image of a typical realisation, with the lower figure showing just the cosmic web emission with logarithmic scaling. The phrase `cosmic web' has been evoked to describe the large scale structure of the Universe as a kind of sponge, and the synchrotron component has often been expected to follow the same hierarchy of structures with sheets, filaments and dense clusters. However, this is not what we observe. Instead, we observe ubiquitous relic-like, shocked shells of emission that surround dark matter halos. Indeed, these are the primary source of emission in the volume. These shocked shells envelop dark matter halos and in the lowest redshift snapshot, for example, $\sim$96\% of the total power in the volume is located in the spherical shell ($r < 1.5 \cdot r_{200}$) of the 100 most massive dark matter halos. The connecting filaments or inflows are not easily discernible, even in a logarithmically scaled image.

In \autoref{fig:fluxradius} we show the distribution of power for the $z = 0.025$ snapshot by plotting the cumulative radio luminosity contained within the spherical volume surrounding each of the 100 most massive dark matter halos, as a function of radius. The cumulative luminosity is plotted in grey for each of the halos, and in blue we plot the mean cumulative luminosity. The universal absence of emission at the core reminds us that in this simulation we do not model Fermi II shock processes such as radio halos. Instead, in general, we observe DSA processes occurring and most luminious in the range $0.75 \cdot r_{200} < r < 1.5 \cdot r_{200}$, although with significant spread in this range from as low as $0.25 \cdot r_{200}$ out to greater than $2 \cdot r_{200}$. The shells themselves are far from isotropic, having complex, filamentary structures with knots and shock fronts orders of magnitude more luminous than elsewhere in the shell.

The absence of significant DSA processes in the innermost cores of dark matter halos is primarily due to the increased matter density and corresponding increase to the speed of sound, as well as the proportionally small area of the shock front; shocks in these regions are therefore not effective electron accelerators despite these regions containing the highest density electron population and magnetic field strengths \citep{Vazza2012}. As the shock proceeds significantly far away from the cluster core it loses energy, encounters an increasingly sparse electron environment, and magnetic field strengths decline; thus similarly these outermost environments are ineffective at producing synchrotron emission. In between, however, there exists a sweet spot which maximises the synchrotron output, precisely as we observe in the simulation.

Moreover, the majority of radio power within the volume surrounds just a handful of interacting clusters. In \autoref{fig:massvspower} we plot the power contained within the spheres centred on dark matter halos with radius $r = 1.5 \cdot r_{200}$ as a function of halo mass. We observe a power law correlation between mass and the cumulative radio power ($L \propto M^{3.6}$; dashed grey line), although we observe significant scattering around this trend that shows the dependence on the specific interaction and merger histories of individual clusters. The plot also makes it clear that just handful of dark matter halo environments account for the majority of the power output in the volume: 90\% of total power is located within the spheres of radius $r = 1.5 \cdot r_{200}$ surrounding just 12 dark matter halos, which cumulatively account for 0.45\% of the total volume.

\subsection{A brief survey of the brightest, most detectable features}

\begin{figure*}
    \centering
    \caption{A sample of emission features at \SI{900}{\mega \hertz} from from various realizations showing both familiar relic formations as well as more unusual shock morphologies. Each image has been convolved to a resolution of \SI{20}{\arcsecond} (beam size indicated by white circle in top right), and the color map [ \SI{}{\jansky \per \beam}] is varyingly scaled from \SI{0}{\jansky \per \beam} to the value of the 99.5th percentile pixel. Red circles indicate the $r_{200}$ dark matter halo extent for halos with $M_{200} >$ \SI{E12}{\solarmass}. Note that all emission occurs outside the core region of dark matter halos, and only appears to be centrally located due to projection effects.}
    \label{fig:interesting}
    \begin{subfigure}{0.32\linewidth}
        \includegraphics[width=\linewidth]{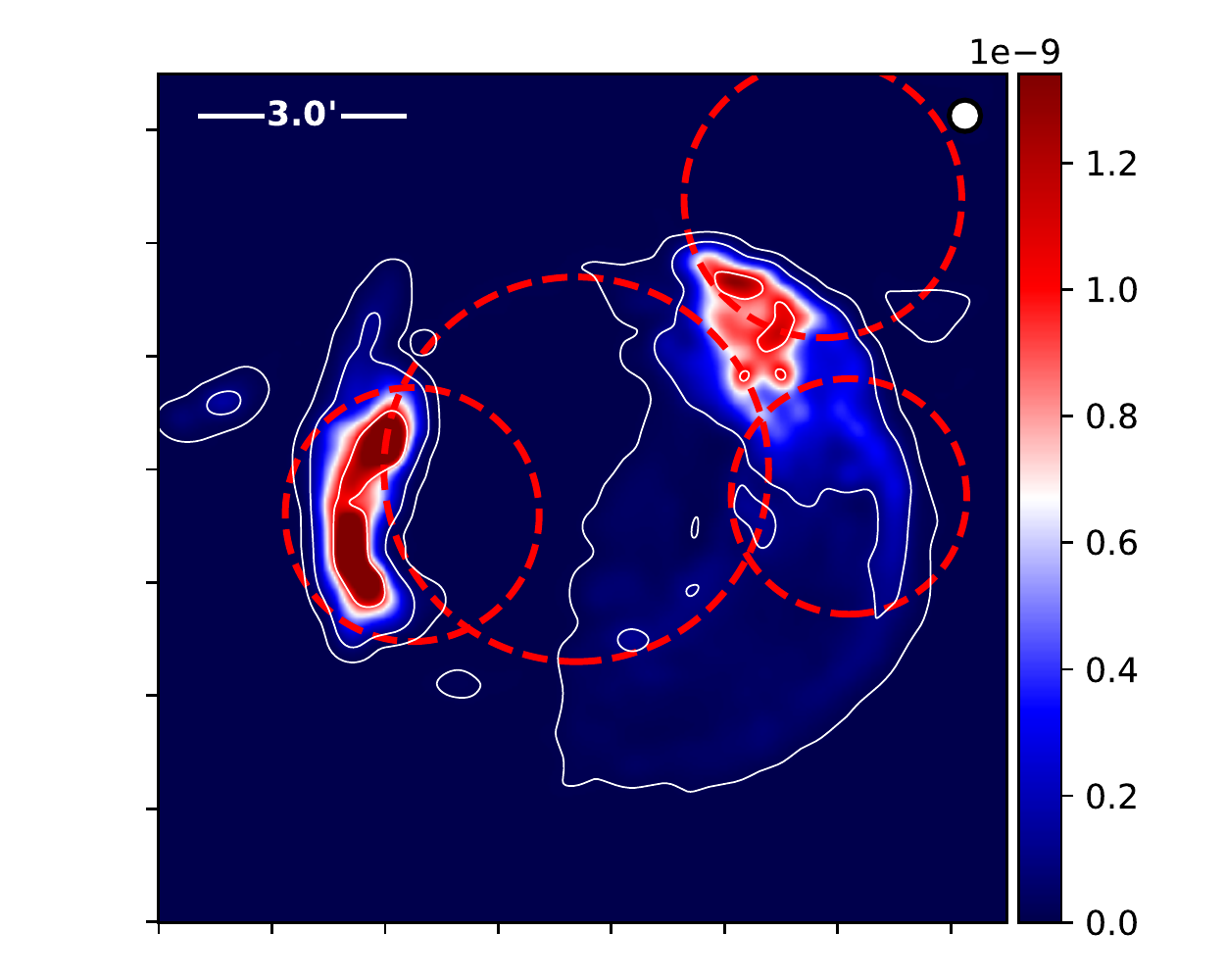}
        \caption{$0.10 < z < 0.15$}
        \label{fig:interesting-2}
    \end{subfigure}
    \begin{subfigure}{0.32\linewidth}
        \includegraphics[width=\linewidth]{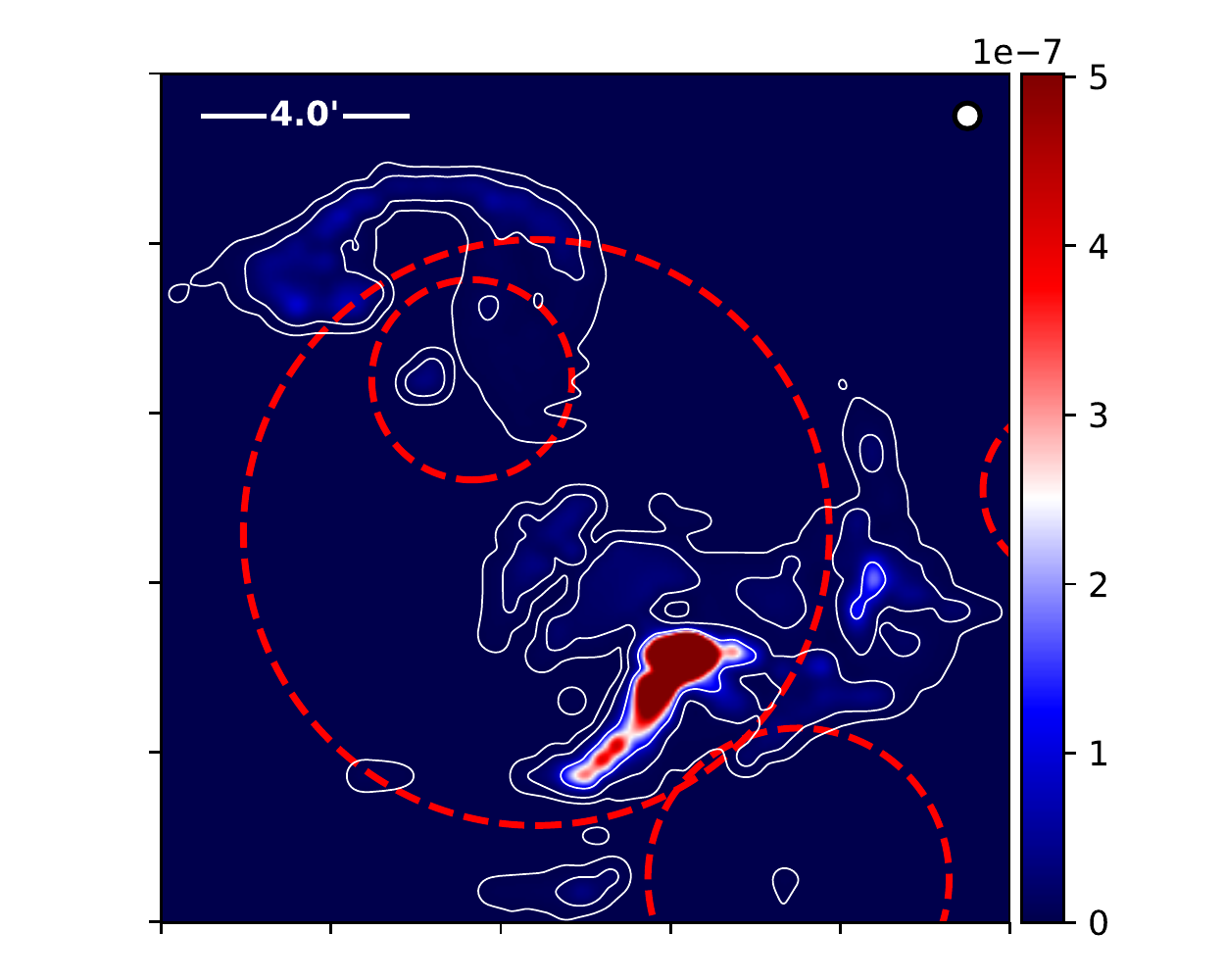}
        \caption{$0.10 < z < 0.15$}
        \label{fig:interesting-3}
    \end{subfigure}
    \begin{subfigure}{0.32\linewidth}
        \includegraphics[width=\linewidth]{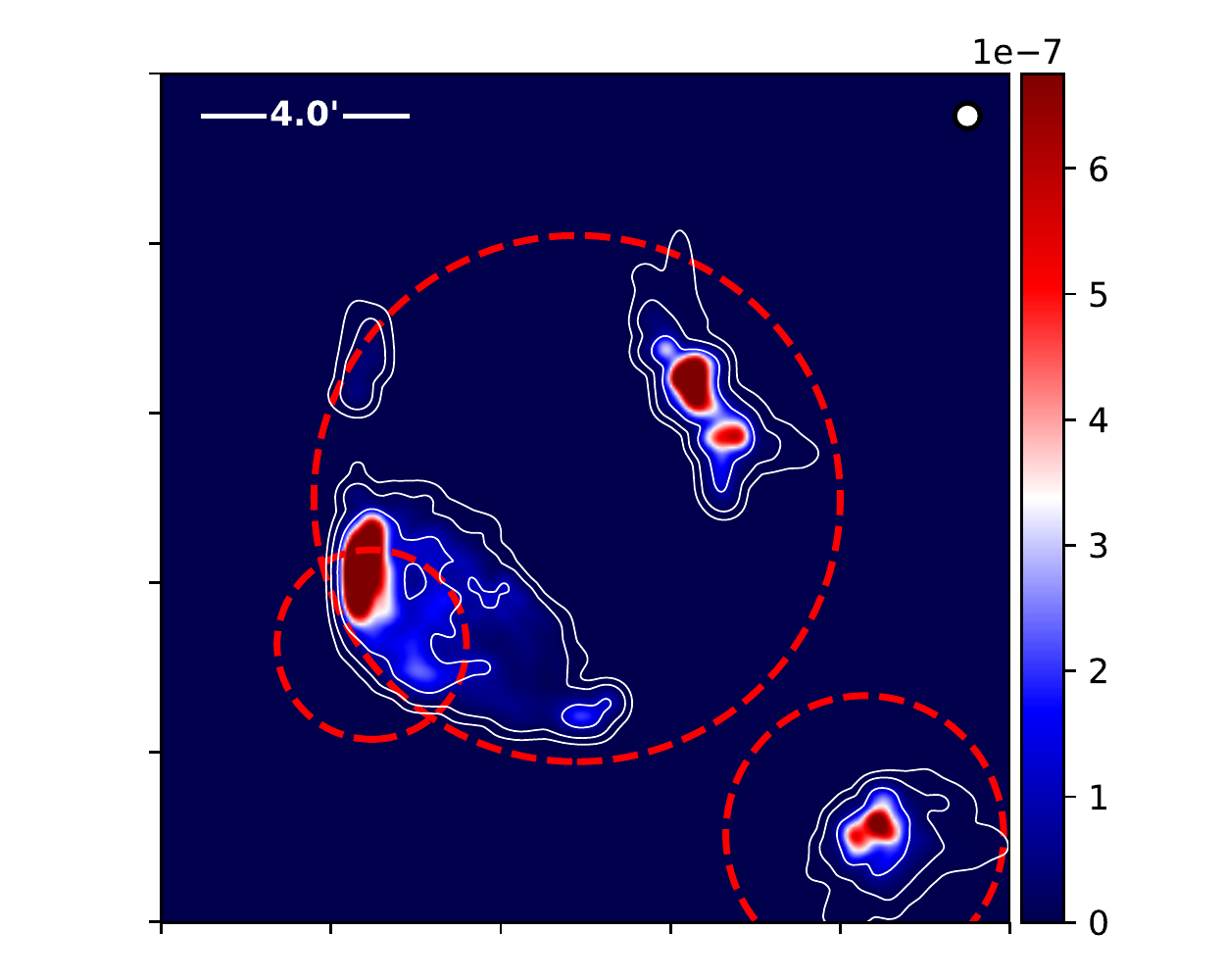}
        \caption{$0.15 < z < 0.2$}
        \label{fig:interesting-5}
    \end{subfigure}
    \begin{subfigure}{0.32\linewidth}
        \includegraphics[width=\linewidth]{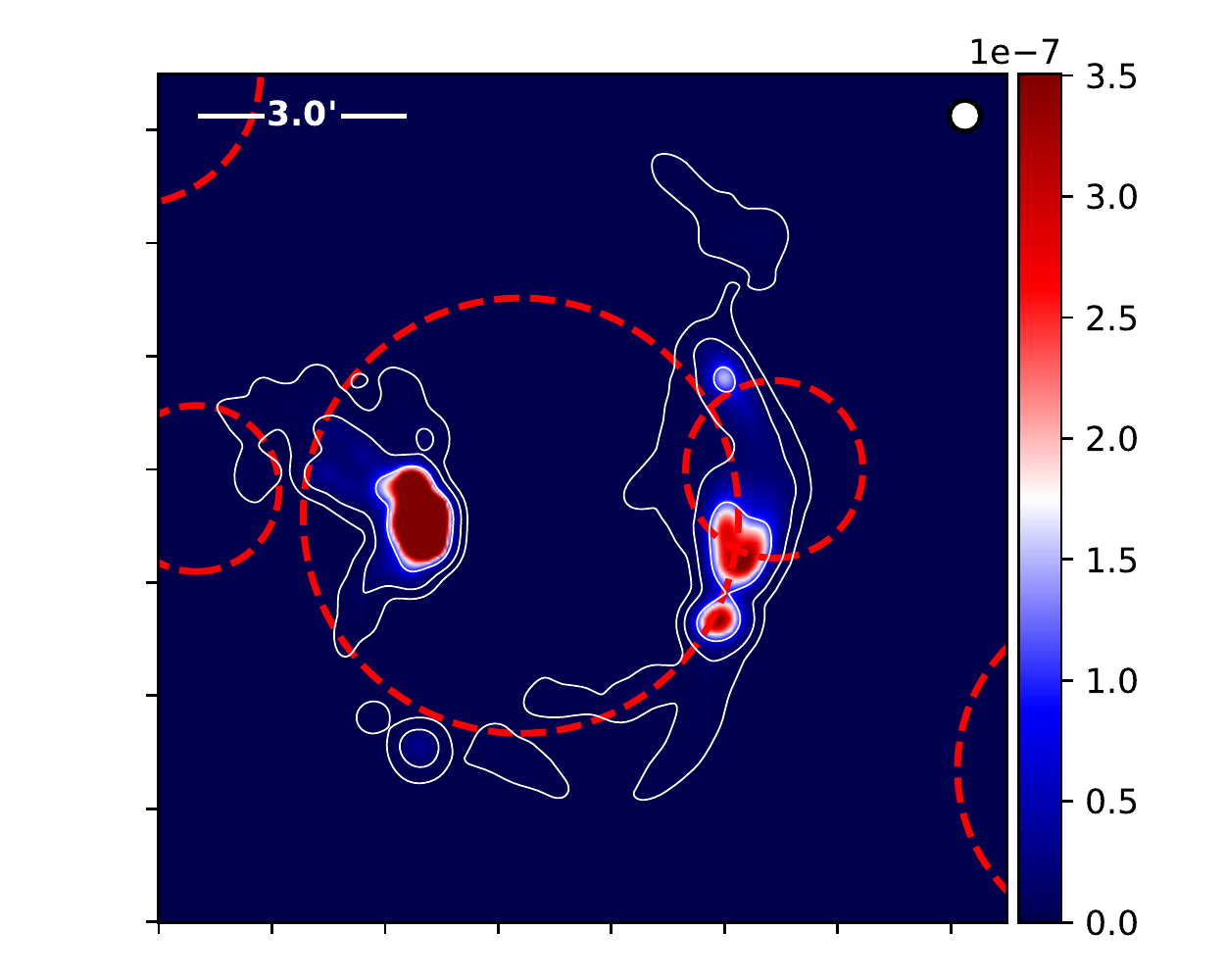}
        \caption{$0.15 < z < 0.2$}
        \label{fig:interesting-7}
    \end{subfigure}
    \begin{subfigure}{0.32\linewidth}
        \includegraphics[width=\linewidth]{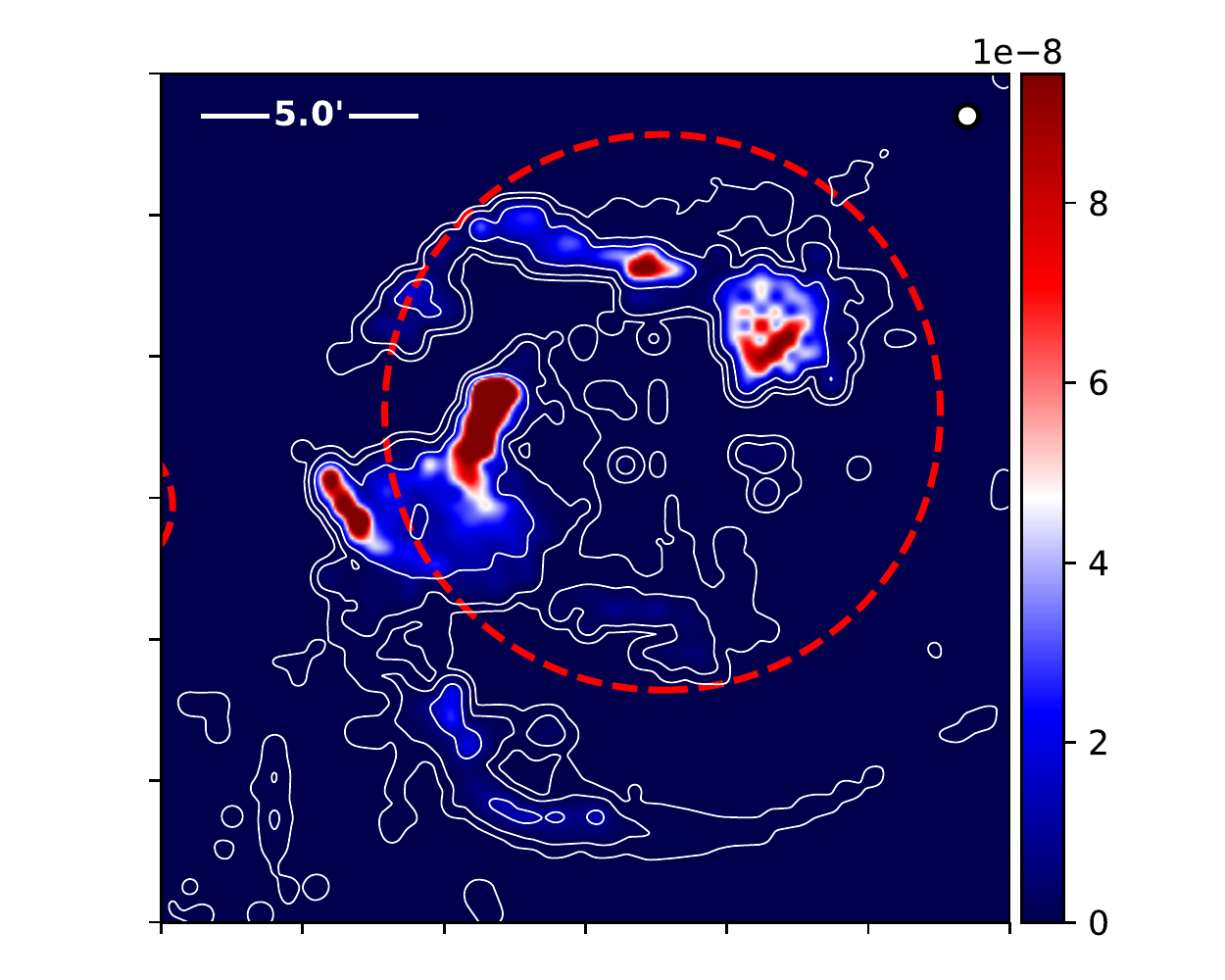}
        \caption{$0.10 < z < 0.15$}
        \label{fig:interesting-4}
    \end{subfigure}
    \begin{subfigure}{0.32\linewidth}
        \includegraphics[width=\linewidth]{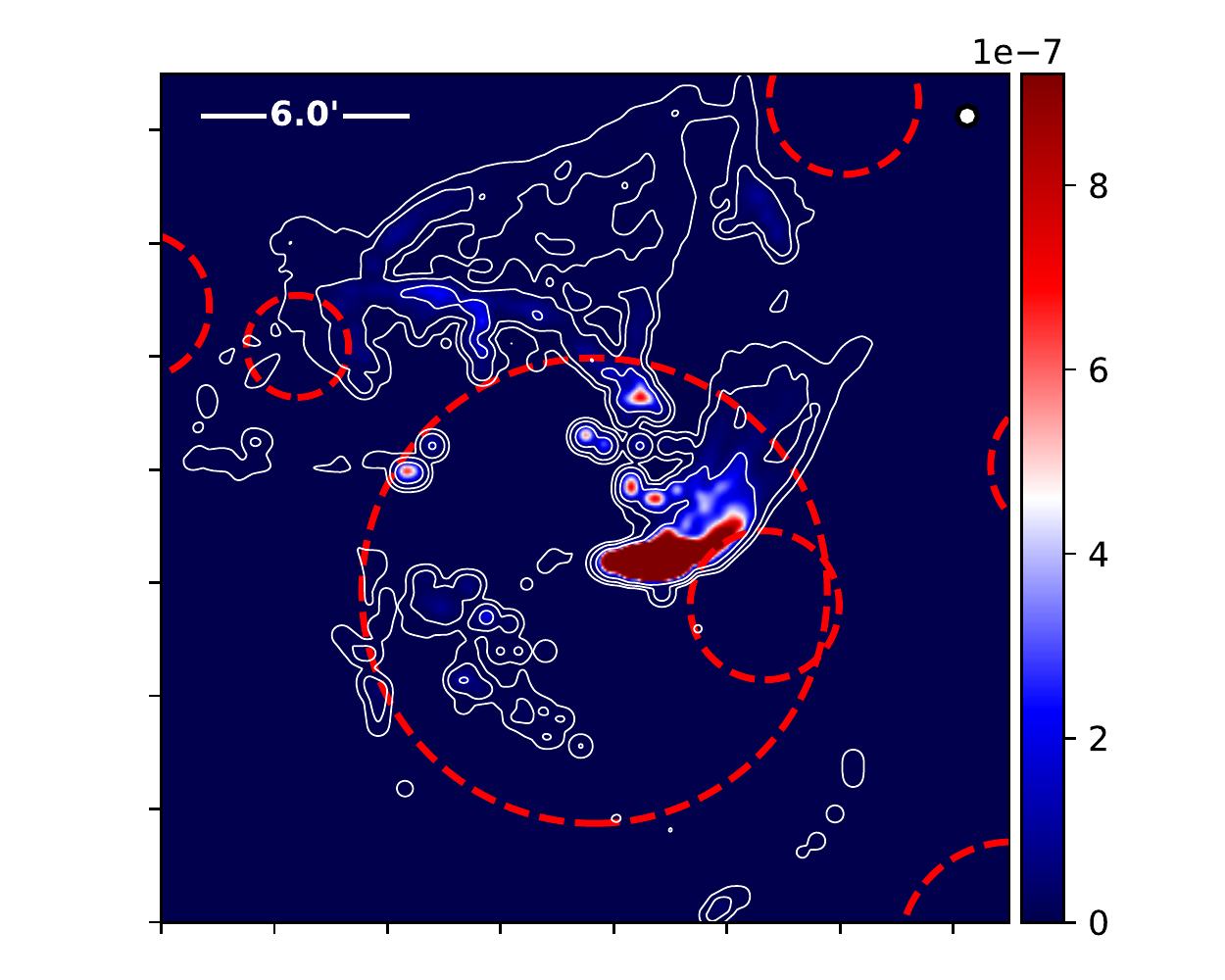}
        \caption{$0.10 < z < 0.15$}
        \label{fig:interesting-9}
    \end{subfigure}
    \begin{subfigure}{0.32\linewidth}
        \includegraphics[width=\linewidth]{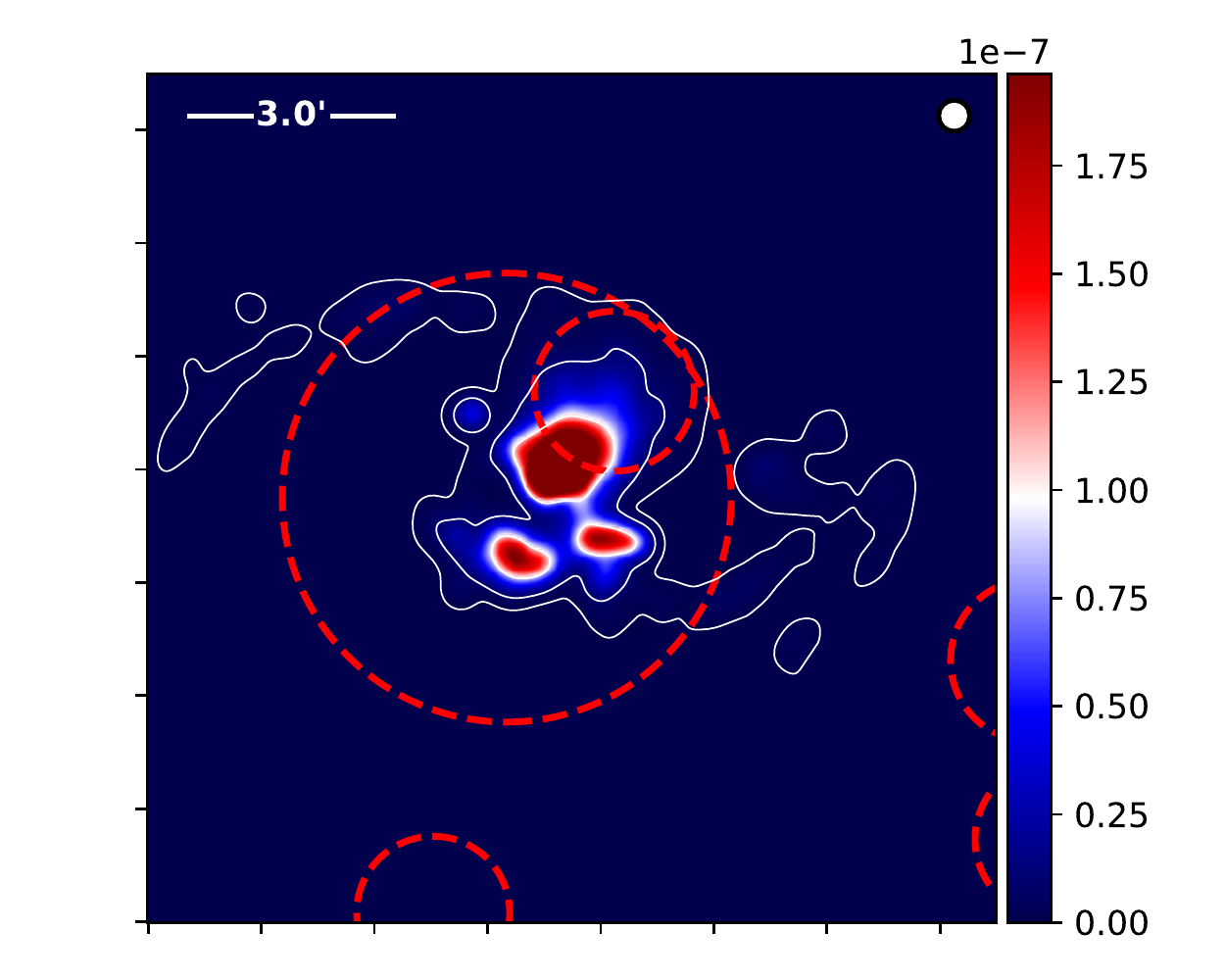}
        \caption{$0.10 < z < 0.15$}
        \label{fig:interesting-8}
    \end{subfigure}
    \begin{subfigure}{0.32\linewidth}
        \includegraphics[width=\linewidth]{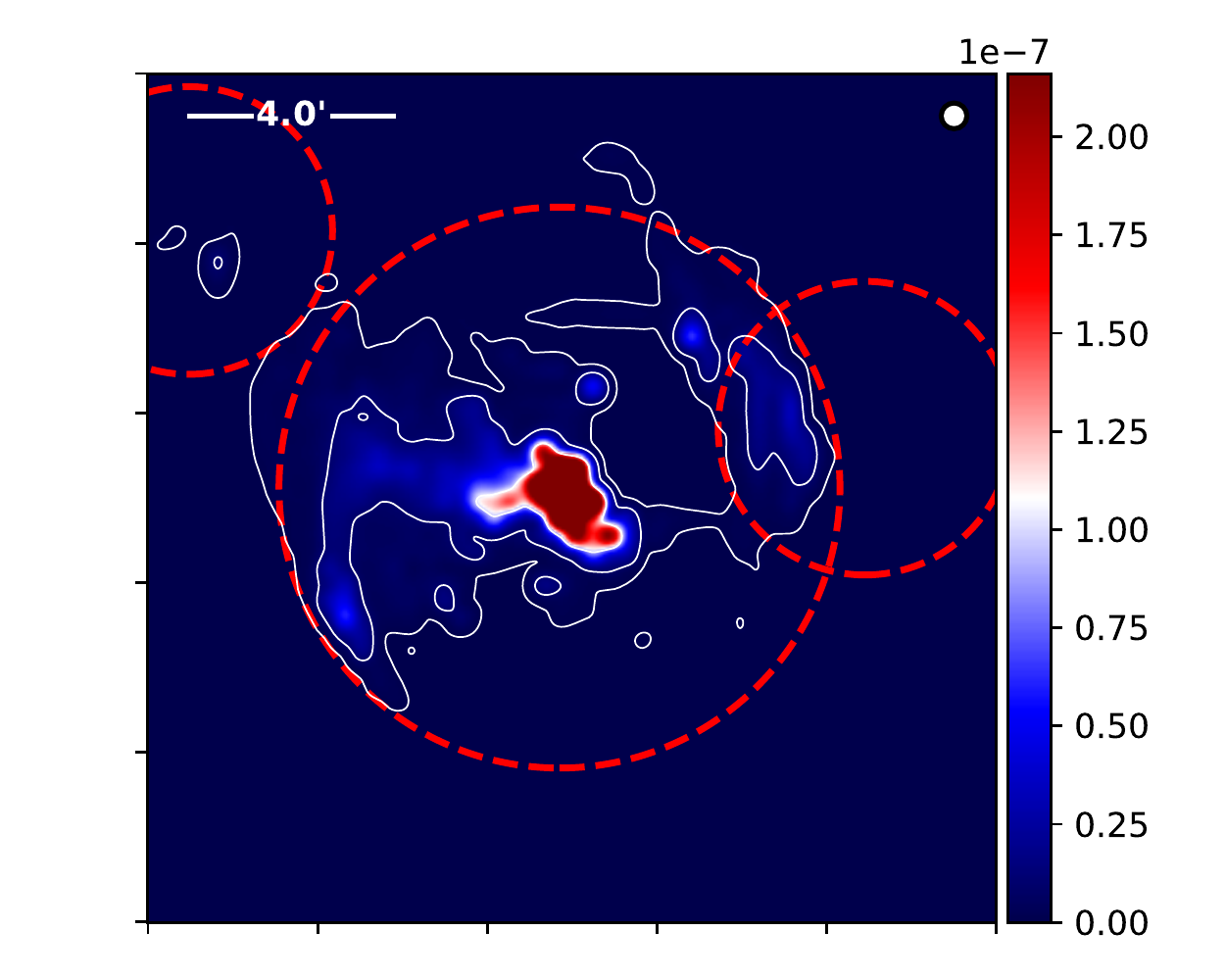}
        \caption{$0.10 < z < 0.15$}
        \label{fig:interesting-10}
    \end{subfigure}
    \begin{subfigure}{0.32\linewidth}
        \includegraphics[width=\linewidth]{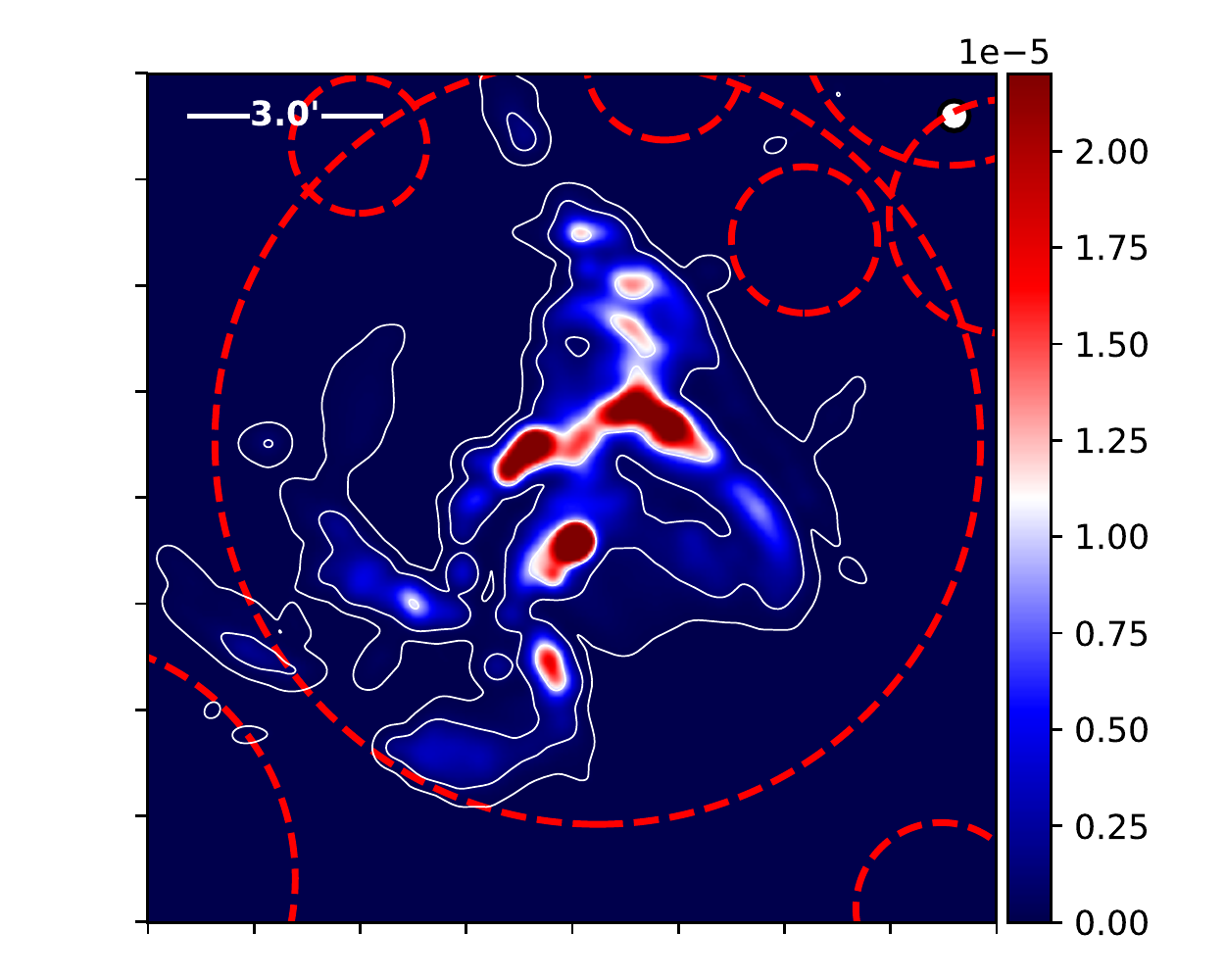}
        \caption{$0.15 < z < 0.2$}
        \label{fig:interesting-1}
    \end{subfigure}
    \begin{subfigure}{0.32\linewidth}
        \includegraphics[width=\linewidth]{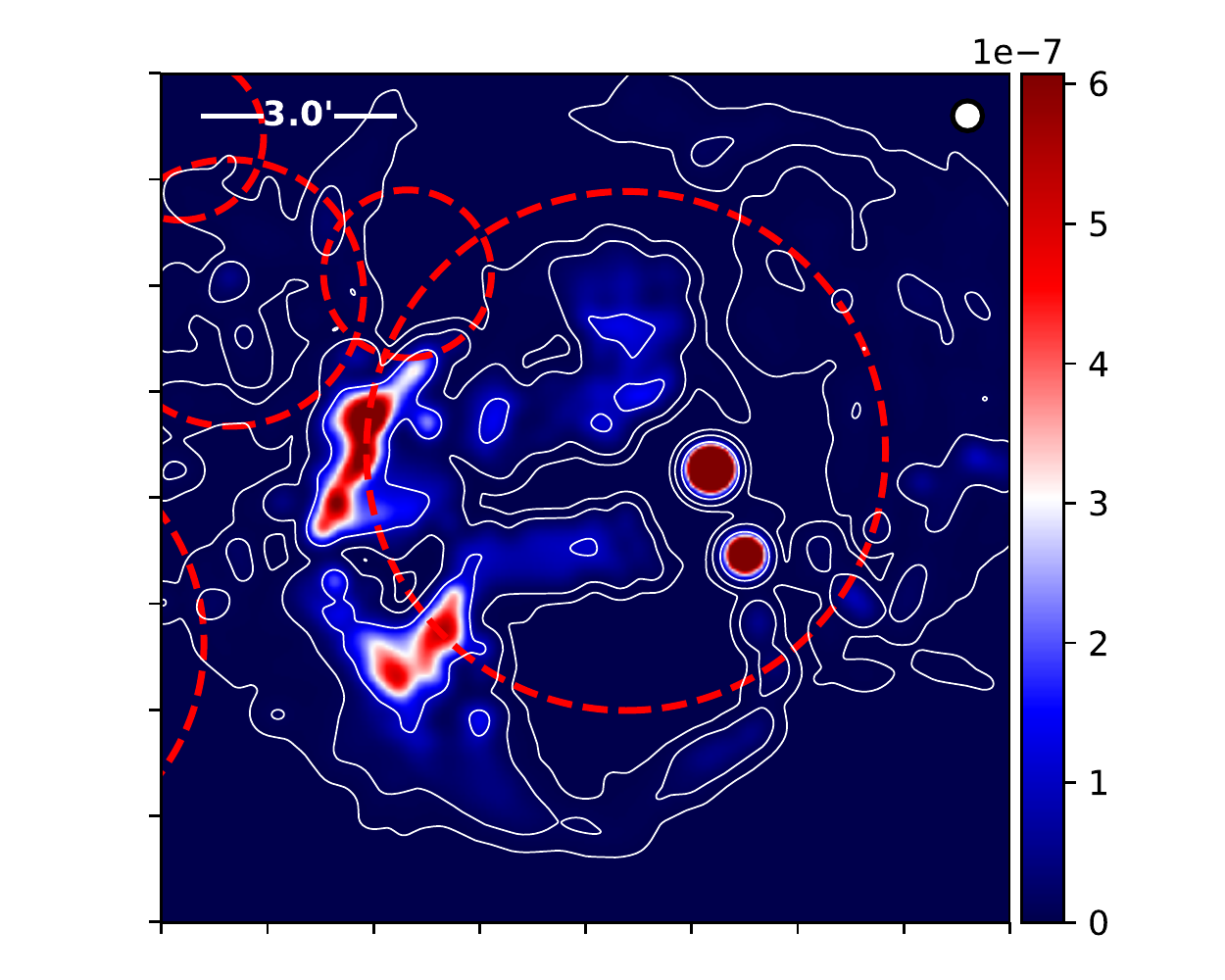}
        \caption{$0.15 < z < 0.2$}
        \label{fig:interesting-12}
    \end{subfigure}
    \begin{subfigure}{0.32\linewidth}
        \includegraphics[width=\linewidth]{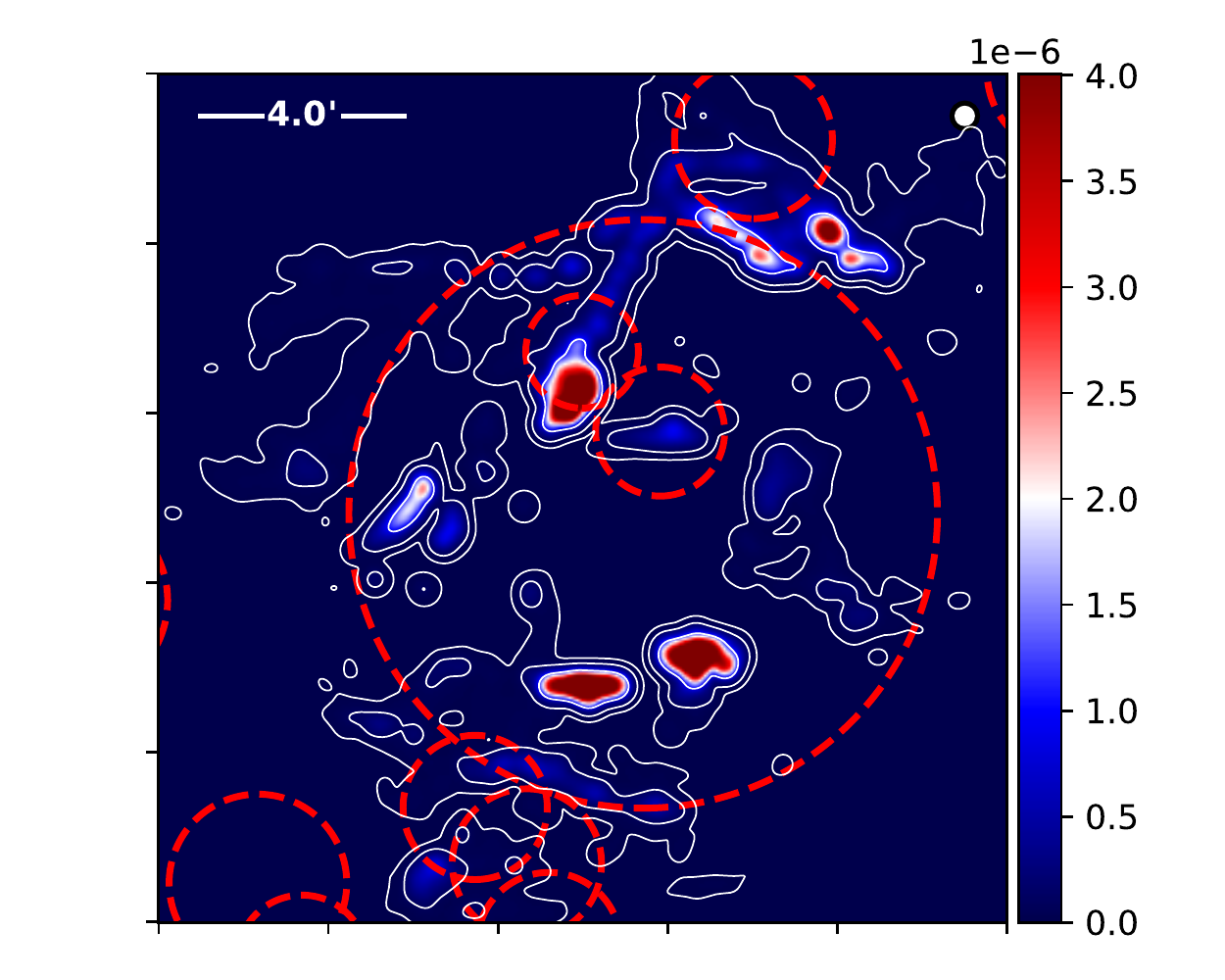}
        \caption{$0.15 < z < 0.2$}
        \label{fig:interesting-6}
    \end{subfigure}
    \begin{subfigure}{0.32\linewidth}
        \includegraphics[width=\linewidth]{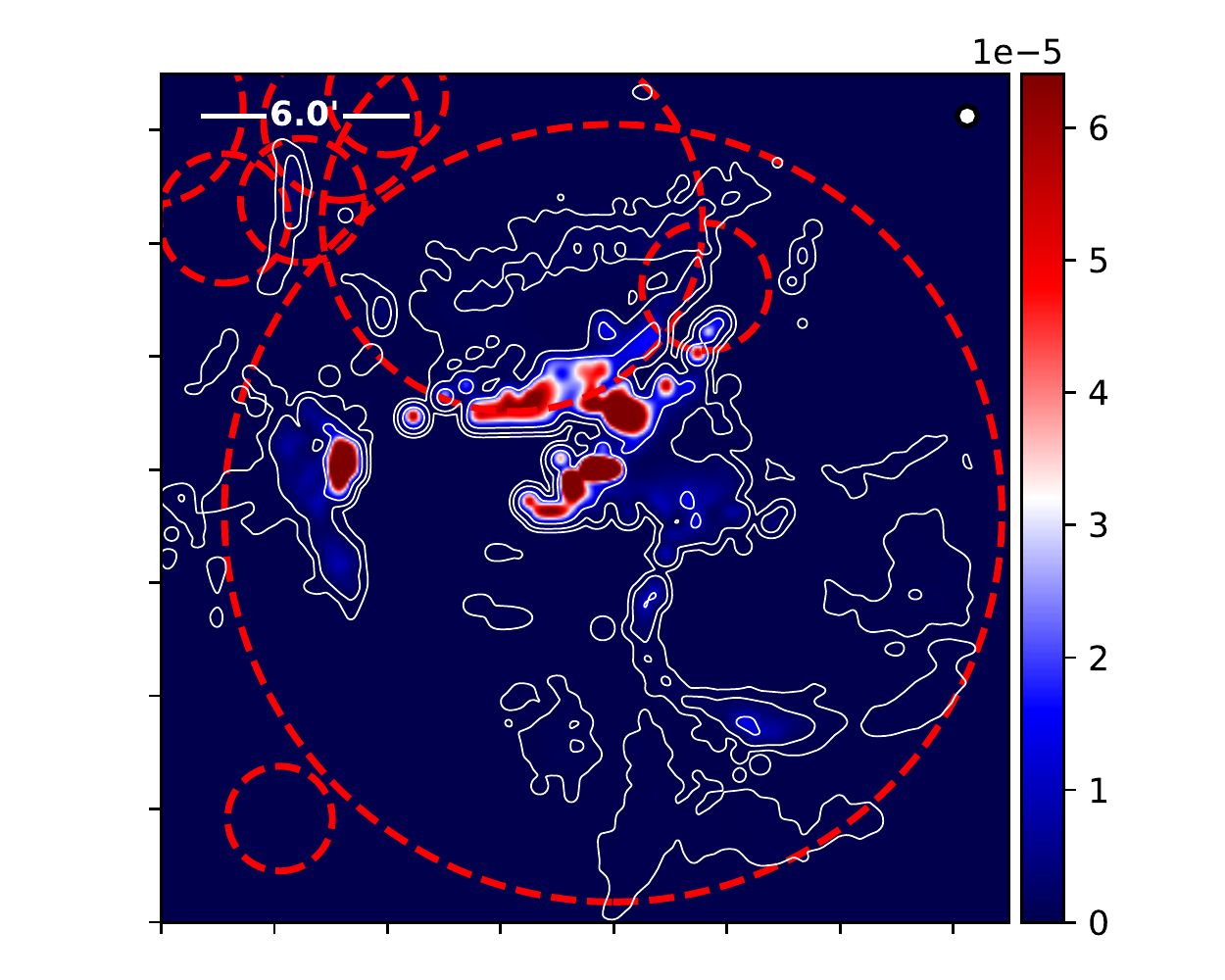}
        \caption{$0.10 < z < 0.15$}
        \label{fig:interesting-11}
    \end{subfigure}

\end{figure*}

To further illustrate the typical emission morphologies, in \autoref{fig:interesting} we provide a small sample of emission regions drawn from redshift slices $0.10 < z < 0.15$ and $0.15 < z < 0.20$ of realizations 1 and 2, observed at \SI{900}{\mega \hertz} and with a \SI{20}{\arcsecond} resolution. These sources were chosen as they are among the brightest emission and provided a good range of morphologies that are present throughout the catalogue. In contrast to previous figures, these are presented using a linearly-scaled colour map, ranging from 0 to the 99.5th percentile pixel, to make clear the morphology and angular size of the emission structures as would actually be observed. This is to correct for the impression of broad, diffuse emission that may be taken from logarithmic color scales, when in fact the emission landscape we observe is one with bright `knotty' peaks of emission and otherwise extremely faint surrounding islands of emission. The contours, however, are logarithmically scaled, indicating 0.1, 0.01 and 0.001 of the 99.5th percentile value. The $r_{200}$ radius of nearby dark matter halos are indicated by red dashed circles.

The most recognisable features in this collection are the very traditional double relic morphologies, which can be seen in figures \ref{fig:interesting-2}, \ref{fig:interesting-3}, \ref{fig:interesting-5}, and \ref{fig:interesting-7}. These examples clearly show matching pairs of arced emission structures as shocked waves travel outwards from the dark matter halo centre, with very classic bow wave morphology. Figure \ref{fig:interesting-4} is more complex but still recognisable as a relic system, whilst \ref{fig:interesting-9} is asymmetric and appears as single relic system.

The remaining examples demonstrate the effect of projection angle on what are otherwise similarly structured shells of emission about dark matter halos. Figures \ref{fig:interesting-8}, \ref{fig:interesting-10} and \ref{fig:interesting-1}, for example, appear in projection to have centrally located emission. Figure \ref{fig:interesting-12} has two apparently point-like peaks of emission that are among the brightest features in this collection, with the brightest peaking at \SI{8}{\micro \jansky \per \beam}; these `points' are aided in their apparent brightness by summing long, narrow emission structures in the radial direction. 

Figures \ref{fig:interesting-6} and \ref{fig:interesting-11} are the most complex in this set of examples, and both involve numerous interacting dark matter halo systems. They are also the most luminous, presumably aided by these cluster interactions. The two Southern peaks in Figure~\ref{fig:interesting-6} peak at approximately \SI{12}{\micro \jansky \per \beam}, whilst the central peak of emission in Figure~\ref{fig:interesting-11} measures \SI{52}{\micro \jansky \per \beam}.

Crucially, note the absence of broad emission features, connecting filamentary bridges between dark matter halos, and in general, emission that isn't associated with shocked shells surrounding dark matter halos.

\subsection{The embedded radio population}

\begin{figure*}
    \centering
    \includegraphics[width=\linewidth,trim={0.45cm 0 0.45cm 0},clip]{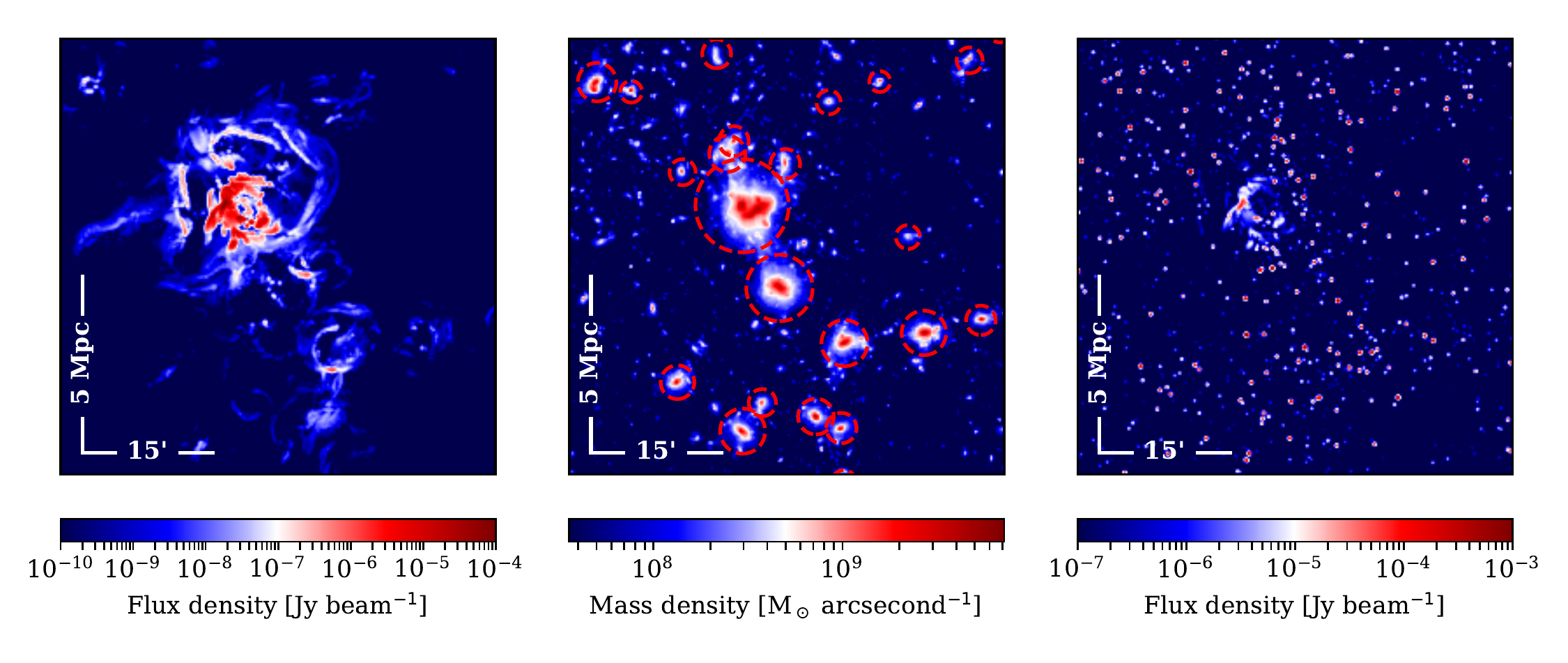}
    \caption{A \SI{50 x 50}{\arcminute} field of view showing redshift range $0.15 \leq z < 0.2$ extracted from realization 3. This redshift range has a comoving depth of \SI{203}{\mega \parsec}, and so this extraction incorporates the full simulation volume stacked twice in the radial direction. The \SI{5}{\mega \parsec} scale has been calculated at the mean redshift $z = 0.175$. \textit{Left:} The synchrotron cosmic web emission at \SI{900}{\mega \hertz} with resolution \SI{20}{\arcsecond}. \textit{Middle:} The associated mass distribution with halos of mass $M >$ \SI{E12.5}{\solarmass} indicated by dashed red circles of radii $r_{200}$. \textit{Right:} The combined cosmic web emission and TRECS radio population for this redshift range at 900 MHz with resolution \SI{20}{\arcsecond}. The TRECS radio population are modelled as simple point sources.} 
    \label{fig:catalogue}
\end{figure*}

We now consider the addition of the T-RECS sources. In \autoref{fig:catalogue} we show a \SI{50 x 50}{\arcminute} field of view showing the redshift range $0.15 <= z < 0.2$ extracted from realization 3 and centred on a massive cluster system. This redshift range has a comoving depth of \SI{203}{\mega \parsec}, and so this extraction incorporates the full simulation volume stacked twice in the radial direction. The panels allow us to compare the distribution of the synchrotron cosmic web in comparison to the underlying mass distribution as well as the embedded radio population.

The left panel shows the cosmic web emission component of this redshift range observed at \SI{900}{\mega \hertz} and at a resolution of \SI{20}{\arcsecond}. The colour scale in the image spans seven orders of magnitiude from the plausibly detectable \SI{10}{\micro \jansky \per \beam} at the heart of the most massive dark matter halo in the field of view, to \SI{10}{\nano \jansky \per \beam} on the virial periphery, to under \SI{1}{\nano \jansky \per \beam} along the filaments. Around the most massive cluster, we observe a number of shocked shells of radio emission; these surround the cluster core, however projection effects mean the brightest emission appears to align with the core region of the cluster.

The central panel shows the mass distribution within the field of view, where we have also overlaid the dark matter halos detected by our halo finding algorithm with mass $M >$ \SI{E12.5}{\solarmass}. We can observe that this is in fact a merging system of two massive systems at the centre of the field, and it is clear that a bridge of increased mass density extends between the two systems. As we observed prior, the northern, most massive system is enshrouded by shocked shells of cosmic web emission, whilst the southern is comparatively quiet in this regard, and there is no significant emission associated with the bridge.

In the rightmost panel of \autoref{fig:catalogue} we show the associated T-RECS radio populations, with both AGN and SFG modelled as simple point sources. The T-RECS sources are not uniformly distributed and display clustering coincident with the most massive portions of the field of view, as well as less dense regions coincident with the dark matter voids. A number of bright T-RECS sources overlap with the peak of the cosmic web emission. Whilst we have modelled the T-RECS sources as simple point sources, at this redshift and at a resolution of \SI{20}{\arcsecond} the lobes of radio loud AGN can be resolved. In this sense, the rightmost panel represents a best case view of the cosmic web, whereas a more sophisticated representation of the T-RECS catalogue would likely occlude the underlying cosmic web emission much more significantly.

\section{Discussion}
\label{sec:discussion}

\begin{table}[]
    \centering
    \begin{tabular}{llll} \toprule
        \textbf{Name} & \textbf{Frequency} & \textbf{Resolution} & \textbf{Noise} \\ \midrule
        MWA & \SI{150}{\mega \hertz} & \SI{60}{\arcsecond} & \SI{1}{\milli \jansky \per \beam} \\
        SKA Low & \SI{150}{\mega \hertz} & \SI{10}{\arcsecond} & \SI{200}{\micro \jansky \per \beam} \\
        ASKAP & \SI{900}{\mega \hertz} & \SI{20}{\arcsecond} & \SI{40}{\micro \jansky \per \beam} \\ \bottomrule  
    \end{tabular}
    \caption{Idealised observing configurations that approximately map to the MWA, SKA Low and ASKAP radio interferometers. The resolution refers to the FWHM of a circular Gaussian beam. }
    \label{tab:instruments}
\end{table}

We orient the discussion of FIGARO primarily around practical questions of the detectability of the cosmic web with current and future radio instruments. This discussion will include its apparent flux distribution, characteristic angular scales, its correlation with the much brighter AGN and SFG populations and, ultimately, the possibility of its detection with the cross-correlation method discussed in the introduction.

To make these comparisons, we will make reference to three idealised instruments that map approximately to: the MWA in its phase 2 configuration \citep[][]{Wayth2018}; the Australian Square Kilometre Array Pathfinder \citep[ASKAP;][]{Hotan2021}; and the proposed Square Kilometre Array Low (SKA Low).\footnote{\url{https://www.skatelescope.org/}} For our purposes, we characterise these instruments simply by an observing frequency, resolution, and noise limit, as shown in \autoref{tab:instruments}.

\subsection{Flux comparison}

\begin{table*}
    \centering
    \begin{tabular}{ccccc} \toprule
        \textbf{Redshift range} & \textbf{T-RECS flux} & \textbf{Web flux} &  \textbf{Web 100th} & \textbf{Web 99.9th} \\
        & [Jy deg$^{-2}$] & [Jy deg$^{-2}$] & [\SI{}{\milli \jansky \per \beam}] & [\SI{}{\milli \jansky \per \beam}] \\ \midrule
        $0 < z < 8$ & 9.06 $\times \left(\nicefrac{\nu}{\text{150 MHz}}\right)^{-0.81}$ & --- & --- & --- \\
        $0 < z < 0.8$ & 3.95 $\times \left(\nicefrac{\nu}{\text{150 MHz}}\right)^{-0.83}$ & 0.027 $\times \left(\nicefrac{\nu}{\text{150 MHz}}\right)^{-1.24}$ & 21 & 0.90 \\
        $0 < z < 0.05$ & 0.064 $\times \left(\nicefrac{\nu}{\text{150 MHz}}\right)^{-0.81}$ & 0.0022 $\times \left(\nicefrac{\nu}{\text{150 MHz}}\right)^{-1.26}$ & 3.9 & 0.12 \\
        $0.05 < z < 0.1$ & 0.12 $\times \left(\nicefrac{\nu}{\text{150 MHz}}\right)^{-0.80}$ & 0.0033 $\times \left(\nicefrac{\nu}{\text{150 MHz}}\right)^{-1.26}$ & 21 & 0.20 \\
        $0.1 < z < 0.15$ & 0.18 $\times \left(\nicefrac{\nu}{\text{150 MHz}}\right)^{-0.84}$ & 0.0034 $\times \left(\nicefrac{\nu}{\text{150 MHz}}\right)^{-1.24}$ & 20 & 0.13 \\
        $0.15 < z < 0.2$ & 0.17 $\times \left(\nicefrac{\nu}{\text{150 MHz}}\right)^{-0.85}$ & 0.0028 $\times \left(\nicefrac{\nu}{\text{150 MHz}}\right)^{-1.22}$ & 11 & 0.12 \\
        $0.2 < z < 0.25$ & 0.21 $\times \left(\nicefrac{\nu}{\text{150 MHz}}\right)^{-0.82}$ & 0.0027 $\times \left(\nicefrac{\nu}{\text{150 MHz}}\right)^{-1.22}$ & 10 & 0.11 \\
        $0.25 < z < 0.3$ & 0.23 $\times \left(\nicefrac{\nu}{\text{150 MHz}}\right)^{-0.82}$ & 0.0021 $\times \left(\nicefrac{\nu}{\text{150 MHz}}\right)^{-1.23}$ & 7.1 & 0.11 \\ \bottomrule
    \end{tabular}
    \caption{Flux statistics across all ten realizations at \SI{150}{\mega \hertz}. For average flux density values, we also provide the flux weighted mean spectral index (as the exponent) allowing for extrapolation up to \SI{1400}{\mega \hertz}. Whilst the flux sums only depend on frequency, the final two columns, the 100th and 99.9th percentile values, are calculated with respect to the idealised MWA configuration.}
    \label{tab:relativefluxstrength}
\end{table*}

\begin{figure*}
    \centering
    \includegraphics[width=\linewidth]{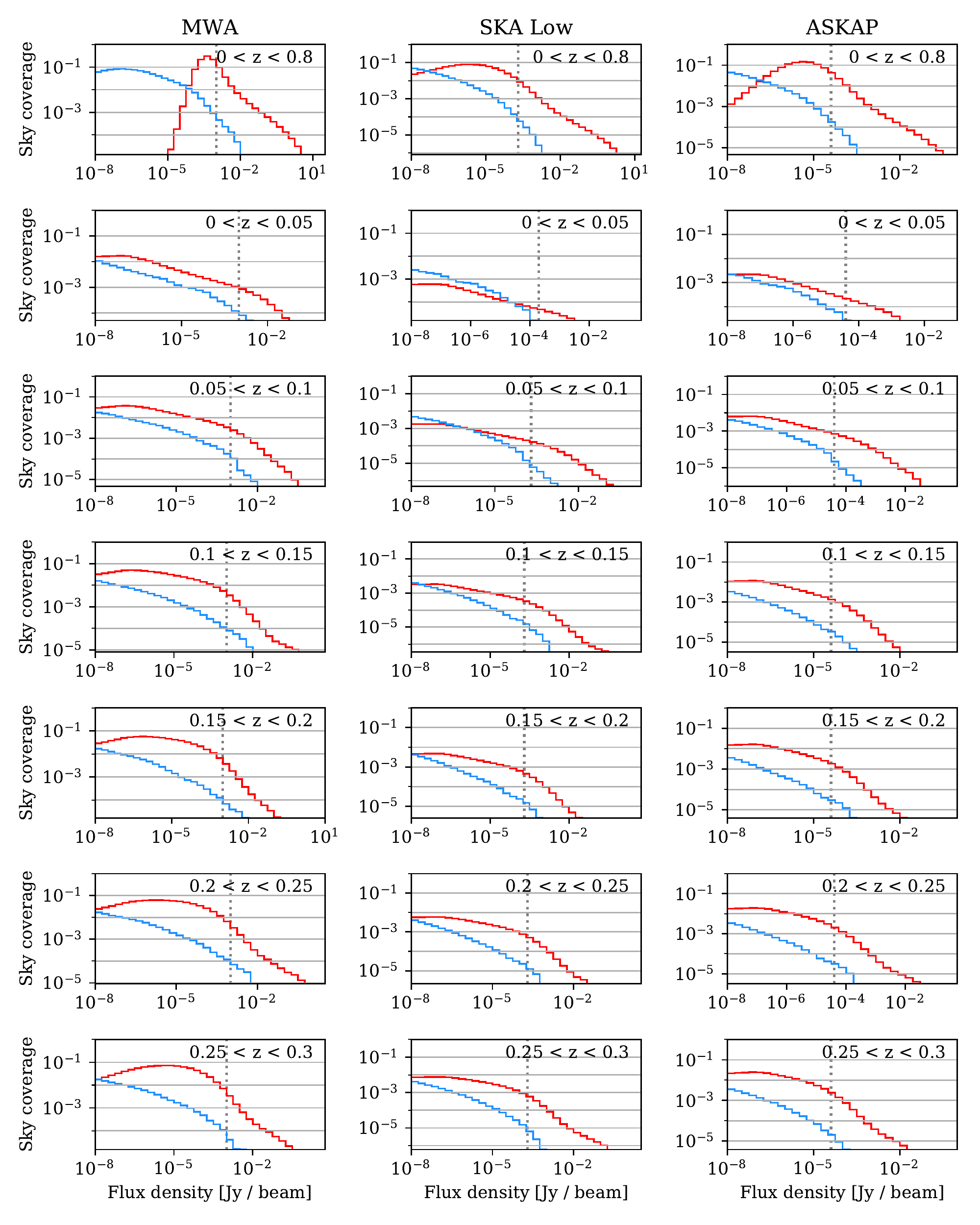}
    \caption{The sky coverage of the cosmic web (blue) and embedded extragalactic (red) emission as a function of flux density binned in $\log_{10}S = 0.25$ increments. We show the sky coverage both as a function of redshift slice, as well as idealised observing configurations for MWA (\SI{150}{\mega \hertz}, \SI{60}{\arcsecond}), SKA Low (\SI{150}{\mega \hertz}, \SI{10}{\arcsecond}), and ASKAP (\SI{900}{\mega \hertz}, \SI{20}{\arcsecond}). Vertical dotted lines indicate the noise threshold for each configuration.}
    \label{fig:fluxdistribution}
\end{figure*}

In this section we quantify flux differences between the embedded radio population and the synchrotron cosmic web. We begin by examining the flux sum across various redshift slices. In \autoref{tab:relativefluxstrength} we show the flux sum at \SI{150}{\mega \hertz} for both of the embedded radio population and the cosmic web emission, computed across all ten realisations. Across the depth of the redshift cone out to $z = 0.8$ the total flux attributed to AGN and SFG sources is \SI{3.95}{\jansky \per \square \deg}, whilst the cosmic web emission over this depth is almost a factor of 150 lower, at \SI{0.027}{\jansky \per \square \deg}. The full flux of the AGN and SFG sources out to $z = 8$ is \SI{8.95}{\jansky}, for which we have no commensurate cosmic web flux values. If we bin the length of the light cones in $\Delta z = 0.05$ slices, we see that the cosmic web emission peaks in the two redshift slices $0.05 < z < 0.1$ and $0.1 < z < 0.15$, each providing a fractional signal of about 0.09\% of the total flux of the simulation.

\autoref{tab:relativefluxstrength} also shows the flux weighted spectral index, which appears as the exponent value, and which allows for extrapolation of the flux sum up to \SI{1400}{\mega \hertz}. These spectral index values do not appear to be redshift dependent across the nearby redshift range we have considered, and show a consistent $\alpha \approx -0.8$ for extragalactic sources and $\alpha \approx -1.25$ for cosmic web emission.\footnote{Since the flux weighted spectral index values of the extragalactic sources are calculated over the frequency range \SIrange[range-phrase=--,range-units=single]{100}{1400}{\mega \hertz}, they will be dominated by the steeper AGN population which are especially luminous at lower frequencies; these values will therefore be steeper than those calculated over higher frequency ranges.} Thus, whilst at \SI{150}{\mega \hertz} the total cosmic web emission is a factor of $\sim150$ times fainter than the embedded extragalactic emission, at \SI{900}{\mega \hertz} this ratio increases to $\sim300$. All things being equal, the cosmic web signal is most detectable at these lower frequencies. It's also important to note that the presence of fossil electron populations, which are not part of our model, may bias the cosmic web emission steeper still.

To further draw out the flux differences between these two populations, we next consider the distribution of flux values on the sky; this is dependent on the observing configuration, and in particular the beam resolution. In \autoref{fig:fluxdistribution} we bin pixels by flux density ($\log_{10}{\Delta S} = 0.25$) and show the proportion of the sky covered by these values, for each of the cosmic web and extragalactic sources when mapped according to the observing configurations in \autoref{tab:instruments}. Vertical dotted lines indicate the noise threshold of each observing configuration. As before, extragalactic sources are modelled simply as point sources. The peak flux distribution of the two populations differs by about 2-3 orders of magnitude for all configurations, with the vast majority of the cosmic web area of emission more than 4-5 orders of magnitude fainter than the bulk of the extragalactic emission. Only a small proportion of the cosmic web is either directly or statistically (e.g.\@ via cross-correlation) detectable. If we refer back to  \autoref{tab:relativefluxstrength}, the final two columns show the peak (i.e.\@ 100th percentile) and 99.9th percentile flux density values when observed in the MWA configuration. The peaks are bright enough that they are directly detectable with the MWA; their morphology, however, tends to resemble the aforementioned apparently point-like knots, making them hard to correctly identify. The extended emission surrounding these knots rapidly declines in brightness, and this can be seen in the 99.9th percentile flux values which are already a factor of $\sim10$ lower than the current best noise limits of the MWA. 

What is not apparent here is the variability of the cosmic web flux distribution between different realizations. In the lowest redshift slice the relatively small volume contained within the \SI{4x4}{\degree}, $0 < z < 0.05$ cone slice results in a highly variable flux distribution which reflects the real degree of cosmic variance of the very nearby universe; thus for this nearest redshift slice, the realizations differ in total flux by 5 orders of magnitude depending on whether they sample from a void, a filament or, by chance, the core region of a massive cluster. By $z > 0.1$, this variability has significantly reduced as the enclosed volume of each redshift slice encompasses greater proportions of the simulation volume; the redshift slice $0.1 < z < 0.15$, for example, differs by just one order of magnitude across all realizations.

\subsection{Cosmic web angular scale}

\begin{figure*}
    \centering
    \includegraphics[width=\linewidth]{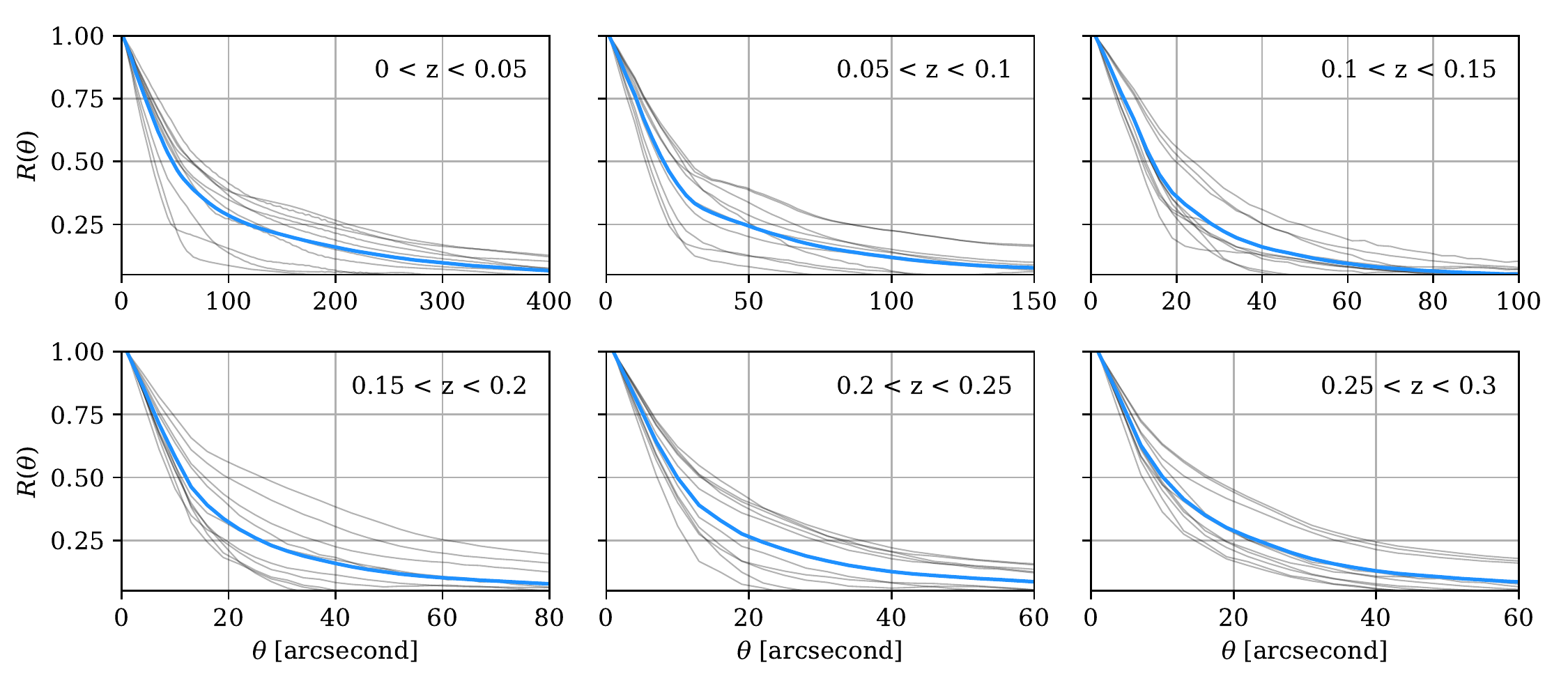}
    \caption{The radial auto-correlation as a function of angular offset, $R(\theta)$, for redshift slices ($\Delta z = 0.05)$ out to redshift $z = 0.3$. The auto-correlation of each specific realisation is shown in grey, and the mean across all ten realisations in shown in blue.}
    \label{fig:auto-correlations}
\end{figure*}

\begin{table}[]
    \centering
    \begin{tabular}{cccc} \toprule
        \textbf{Redshift} & \textbf{min($\theta$)} & \textbf{mean($\theta$)} & \textbf{max($\theta$)} \\
        & [arcsecond] & [arcsecond] & [arcsecond] \\ \midrule
        $0 < z < 0.05$ & 55 & 105 & 149 \\
        $0.05 < z < 0.1$ & 28 & 43 & 58 \\
        $0.1 < z < 0.15$ & 22 & 31 & 48 \\
        $0.15 < z < 0.2$ & 18 & 28 & 53 \\
        $0.2 < z < 0.25$ & 14 & 22 & 31 \\
        $0.25 < z < 0.3$ & 14 & 22 & 33 \\\bottomrule
    \end{tabular}
    \caption{The characteristic angular scale of cosmic web emission, measured here by the FWHM of the auto-correlation of the cosmic web maps, for redshift slices of $\Delta z = 0.05$ out to $z = 0.3$. The minimum, mean and maximum are calculated across each of the ten realisations and are indicative of the expected cosmic variance between \SI{4 x 4}{\degree} fields.}
    \label{tab:auto-correlation}
\end{table}

We next consider the characteristic angular extent of cosmic web sources, quantified through the radial auto-correlation function. This is described in \autoref{appendix:crosscorrelation}, with the modification that we cross-correlate the map against itself (i.e.\@ we set map $A = B$). \autoref{fig:auto-correlations} shows the radial auto-correlation as a function of angular offset, $R(\theta)$, for a number of redshift slices out to redshift $z = 0.3$. This redshift range is typical of previous detection attempts, for example \citet{Brown2011} which only extended as deep as $z = 0.05$ and \citet{Vernstrom2017} which had a mean redshift depth of $z = 0.2$. The auto-correlation of each specific realisation is shown in grey, and the mean across all ten realisations is shown in blue; the large spread of results, especially at low redshifts, is primarily a result of cosmic variance.

To compare results, \autoref{tab:auto-correlation} compiles the minimum, mean and maximum FWHM values for each redshift slice. In the lowest redshift bin $(0 < z < 0.05)$, $\theta_\text{FWHM}$ varies from as small as \SI{55}{\arcsecond} to as much as \SI{149}{\arcsecond}; for redshift bin $(0.05 < z < 0.1)$ the range is \SIrange[range-phrase=--,range-units=single]{28}{58}{\arcsecond}; and for the redshift bin $(0.1 < z < 0.15)$ the range is much more narrow at \SIrange[range-phrase=--,range-units=single]{28}{43}{\arcsecond}. Given that these volumes are much smaller than the simulation volume (refer to \autoref{tab:fractionalvolume}), we believe that these ranges are likely representative of the real spread. The deepest three redshift bins in \autoref{fig:auto-correlations} have mean values for $\theta_\text{FWHM}$ of \SI{28}{\arcsecond}, \SI{22}{\arcsecond} and \SI{22}{\arcsecond}, respectively, however these values are increasingly likely to be affected by systematics arising from the limited simulation volume.

These angular sizes are significantly less than the typical angular length of intracluster filaments at their respective redshifts, and this latter scale has often been used as a proxy for the expected angular extent of cosmic web emission \citep[e.g.][]{Vernstrom2017}. This smaller than expected angular extent arises as a result of the `knotty' morphology of most emission structures, combined with our previous observation that the emission does not, in general, bridge cluster pairs but rather is restricted to halo shells. Whilst previous detection attempts, such as \citet{Vacca2018}, have deliberately chosen observing strategies to ensure sensitivity to extended and diffuse emission structures presumed to have angular scales that are multiple arcminutes in extent, these results suggest that a much finer resolution on the order of \SIrange[range-phrase=--,range-units=single]{20}{40}{\arcsecond} is best for nearby cosmic web emission.

\subsection{Cross-correlating the cosmic web}

\begin{figure}
    \centering
    \begin{subfigure}{1\linewidth}
        \includegraphics[width=\linewidth,clip,trim={0 0.5cm 0 0}]{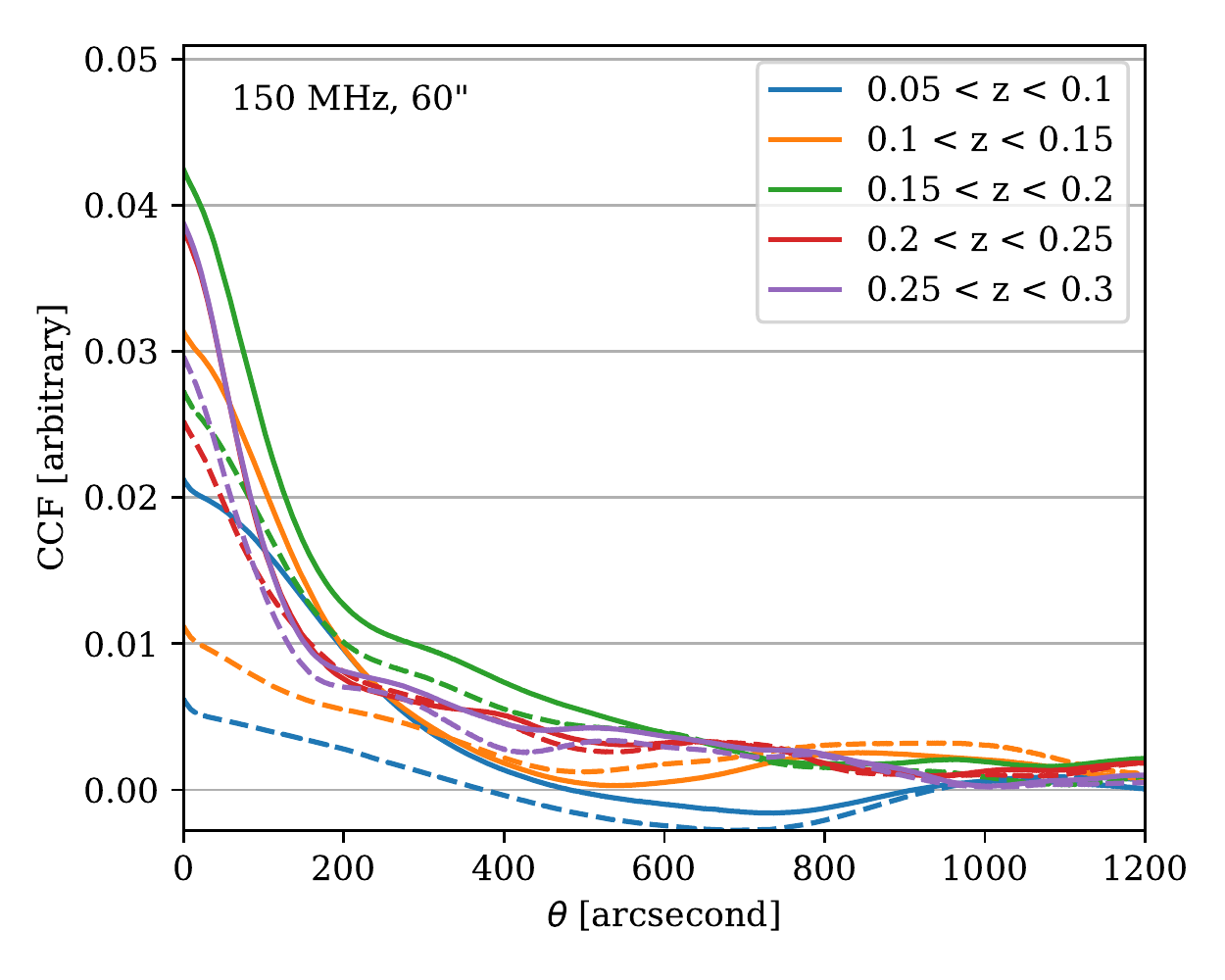}
        \caption{MWA}
    \end{subfigure}
    \begin{subfigure}{1\linewidth}
        \includegraphics[width=\linewidth,clip,trim={0 0.5cm 0 0}]{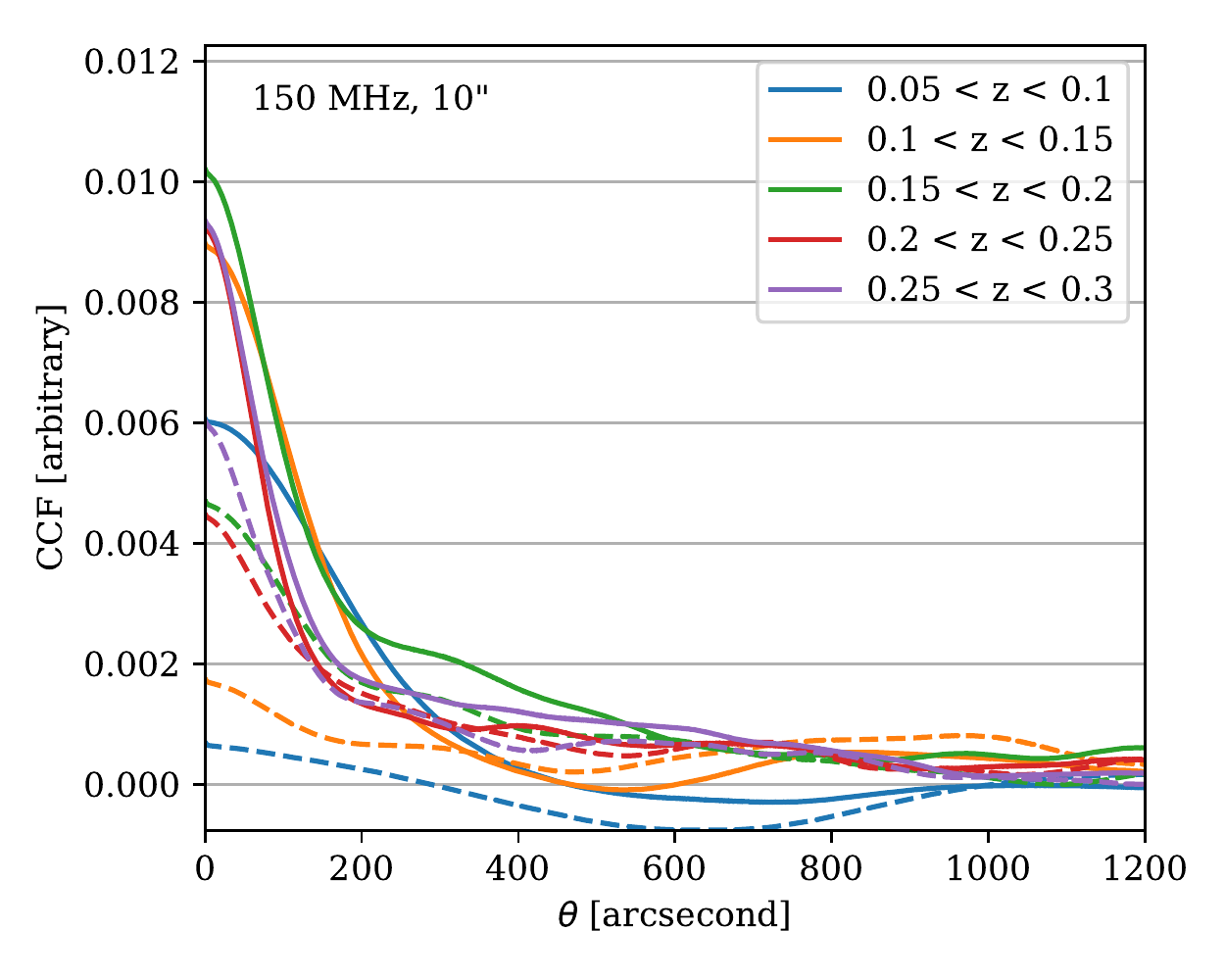}
        \caption{SKA Low}
    \end{subfigure}
    \begin{subfigure}{1\linewidth}
        \includegraphics[width=\linewidth,clip,trim={0 0.5cm 0 0}]{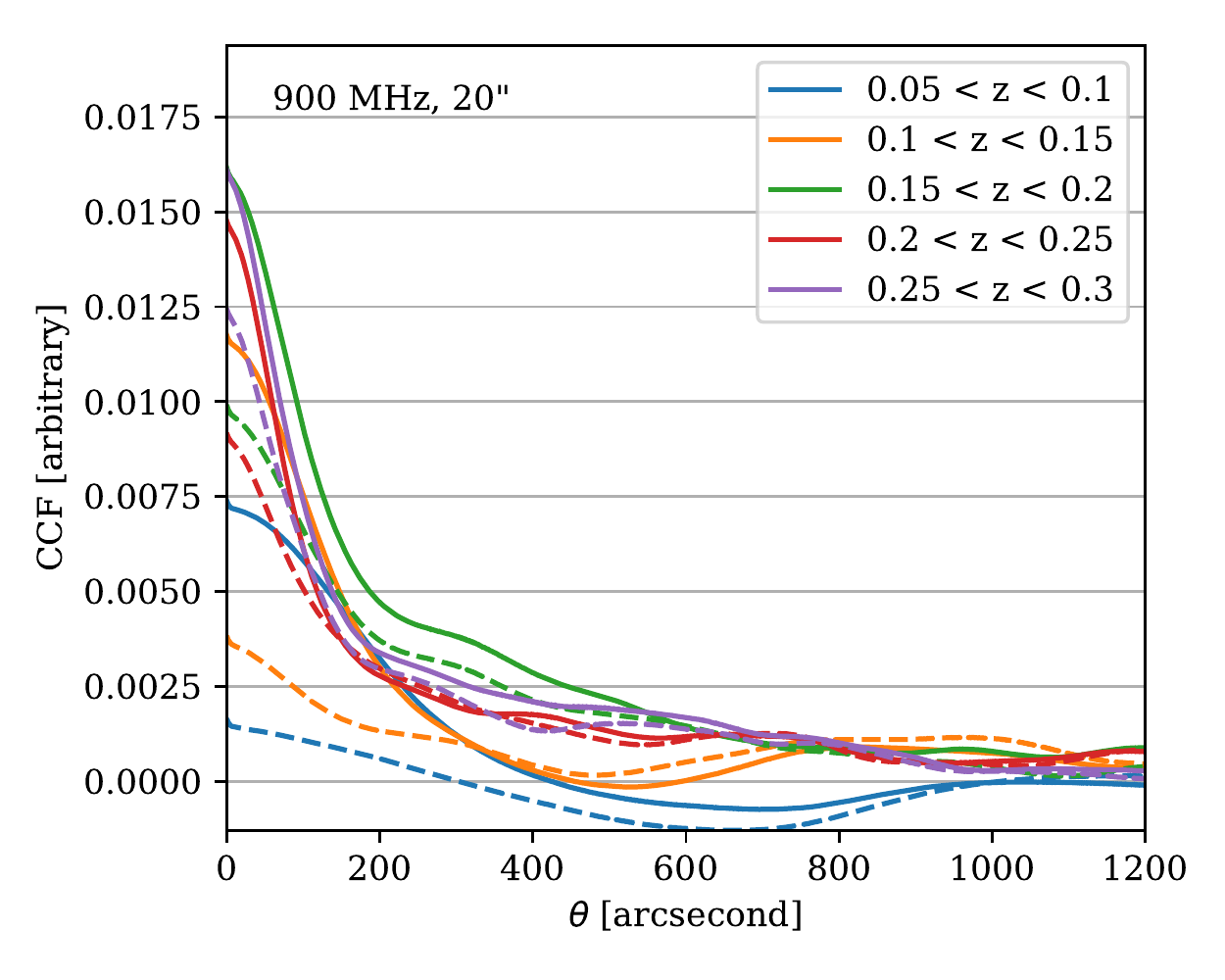}
        \caption{ASKAP}
    \end{subfigure}
    \caption{Cross-correlation of mass density with the FIGARO simulations, for a variety of redshift slices (solid lines), and compared with the `null' case where the underlying cosmic web emission has been flipped (dashed). The excess correlation versus the null result therefore shows the cosmic web component of the cross-correlation result.}
    \label{fig:masscorrelation}
\end{figure}

\begin{figure*}
    \centering
    \includegraphics[width=\linewidth,clip,trim={0cm 0 0cm 0}]{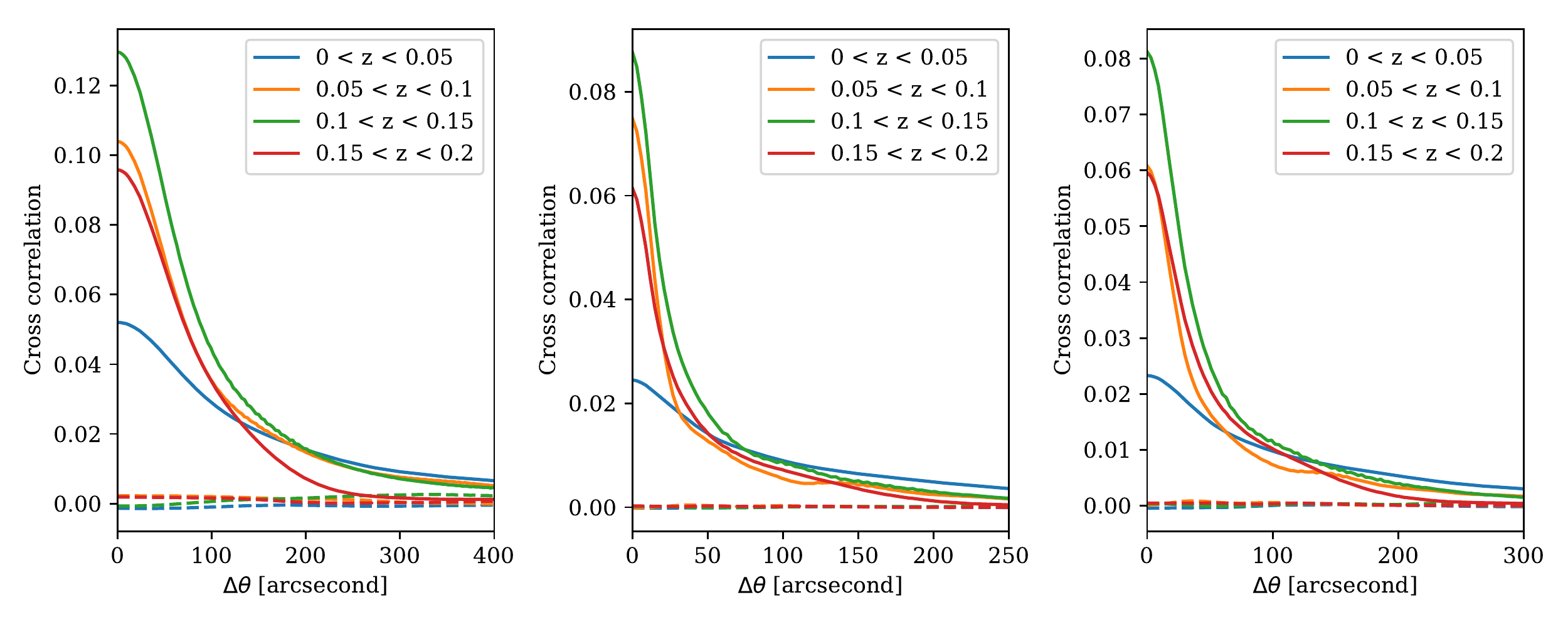}
    \caption{Cross-correlation of the all FIGARO realizations with kernels of the \textit{known} cosmic web signal for a variety of redshift slices (solid line), in comparison with a `null' result where the cosmic web signal has been spatially flipped (dashed line). The cosmic web emission is easily detectable even in amongst the extragalactic sources. Note the lack of correlation peak for the null result, showing the extragalactic sources do not cluster in the same way as the cosmic web emission. \textit{Left:} MWA, \SI{150}{\mega \hertz}, \SI{60}{\arcsecond} resolution. \textit{Middle:} SKA Low, \SI{150}{\mega \hertz}, \SI{10}{\arcsecond} resolution. \textit{Right:} ASKAP, \SI{900}{\mega \hertz}, \SI{20}{\arcsecond} resolution.}
    \label{fig:detectability}
\end{figure*}

The cosmic web signal is too faint to detect directly, with the exception of outlier emission knots, and so one promising method is the radial cross-correlation method used by both \citet{Brown2017} and \citet{Vernstrom2017}. In this method, a kernel image is first constructed and weighted based on where cosmic web emission is believed to concentrate across a map. This `best guess' map for the cosmic web is then cross-correlated and radially averaged in an effort to bring out the cosmic web signal that is otherwise hidden beneath the image noise; a peak at or very near \SI{0}{\degree} offset suggests a possible detection of the cosmic web. In \citet{Vernstrom2017}, the correlation kernel was produced by using galaxy density maps, which were believed to be a proxy for large scale mass density and in turn for the synchrotron component of the cosmic web. A challenge with this approach is that the extragalactic radio population also correlates with this density map, producing an excess cross-correlation signal that is difficult to separate from any potential cosmic web signal.

We here attempt to reproduce this `false' correlation by cross-correlating FIGARO with the mass density maps pertaining to redshift slices of depth $\Delta z = 0.05$, for the redshift range $0.05 < z < 0.3$. Prior to cross-correlation, the mass density maps were smoothed with a Gaussian kernel having a FWHM of \SI{1}{\mega \parsec} pertaining to the mean redshift of the respective slice. In \autoref{fig:masscorrelation} we show the cross-correlation of FIGARO for each of the idealised observing configurations with these mass density maps (solid line), as well as a `null' result where the underlying cosmic web emission has been spatially flipped (dashed line). All AGN and SFG sources have been represented as simple point sources and have been Cleaned down to the noise threshold of the respective observing configuration. We observe that all results peak at zero, indicating a positive correlation with the mass density maps. In the case of the null result (dashed line), this correlation is solely the result of the AGN and SFG populations clustering similarly to the underlying mass distribution; this `false' signal is precisely the issue \citet{Vernstrom2017} encountered. However, we also observe an excess correlation in the solid line, and this is due to the additional presence of the cosmic web. The degree of this excess is quite significant, accounting for more than half of the signal in the closest redshifts, and reducing somewhat at higher redshifts. We should note that the presence of other cluster emission that is not modelled in our simulation, such as radio halos, faint and diffuse AGN remnants, and radio phoenix, would further strengthen the `null' result signal and reduce the relative excess; these populations are not well understood and it is very difficult at this time to model their additional contribution to the cross-correlation signal.

This spatial alignment of unrelated cluster emission and cosmic web emission makes detection extremely challenging. Nonetheless, we can still ask the question: if we had perfect knowledge of the distribution of the cosmic web, can we avoid the cross-correlation from being polluted by extragalactic population? To answer this, we have computed the cross-correlation of the full FIGARO maps (cosmic web, AGN and SFG sources combined) with a kernel that is the cosmic web emission located in webshift slices of depth $\Delta z = 0.05$. We do this for all three observing configurations and have cleaned each map down to its respective noise threshold to produce a residual image. In \autoref{fig:detectability}, we present the results of this cross-correlation and also additionally a `null' result where we have spatially flipped the cosmic web signal in the map. The cosmic web kernel correlates strongly with itself, as expected, even in amongst the additional AGN and SFG residual sources; and moreover, this correlation seems to peak in the redshift slice $0.1 < z < 0.15$ suggesting this is a sweet spot that maximises cosmological volume against the fading brightness that arises with increasing luminosity distance. The strongest correlation is observed in the MWA configuration, which by observing at \SI{150}{\mega \hertz} takes advantage of the steep spectral index of cosmic web emission, but which is also advantaged by its lower resolution beam that is approximately at the characteristic scale of the cosmic web emission. Despite the disadvantage of observing at a higher frequency, the ASKAP configuration also produces a strong correlation signal suggesting this might well be a valid observing configuration.

Most significantly, however, we observe no significant signal in the null results in \autoref{fig:detectability}, and in fact in many cases this signal is not only weak, it is also weakly negative indicating anti-correlation. This is an important result: on average, the cosmic web signal is spatially separate and distinct from the extragalactic radio population, and they do not cluster in the same way. Whilst this provides for the possibility of devising a kernel that does not correlate with other cluster emission, of course any real detection attempt will not be blessed with perfect knowledge of the cosmic web distribution. The much more difficult task will be in devising kernels based only on approximate knowledge and proxies such as the mass distribution, but which do not strongly correlate with central cluster emission. We leave the development of kernels based only on limited knowledge of the Universe to future work.

\section{Conclusion}

We have released the FIGARO simulation: the first combined simulation of the cosmic web and its embedded extragalactic radio population spanning ten unique \SI{4 x 4}{\degree} fields, out to a redshift of $z = 0.8$, and valid over a frequency range \SIrange[range-units=single]{100}{1400}{\mega \hertz}. In doing this, we have brought together the largest MHD simulation to date encompassing $100^3$~Mpc$^3$---from which we have derived a model of synchrotron cosmic web emission, calibrated to match observed radio relic population statistics---and combined this simulation with the previously released T-RECS codebase to model the embedded extragalactic AGN and SFG embedded populations. Uniquely, these latter populations have been positioned such that they follow the underlying mass distribution of the MHD simulation, and thus cluster realistically alongside the associated cosmic web emission. We make these simulations publicly available as we believe they can be helpful in developing cosmic web detection techniques, as well as in modelling and constraining confounding signals that may arise in the course of these detection attempts.

In addition, we have provided an early analysis of the FIGARO simulation and its properties. Foremost amongst these results is to emphasise the spatial distribution and morphology that the cosmic web takes: as opposed to a sponge-like, filamentary structure that traces the underlying mass distribution, we find that the radio cosmic web is composed of ubiquitous, relic-like shells, principally in the spherical annuli $0.75 \cdot r_{200} < r < 1.5 \cdot r_{200}$ enshrouding dark matter halos. The filaments proper are largely empty of any significant emission. Moreover, the actual distribution of power in these shells is irregular and `knotty'. The brightest of these knots rise to the level of detection with the current generation of instruments whilst the surrounding emission rapidly declines in brightness, with a characteristic angular extent of $\lessapprox$\SI{58}{\arcsecond} already by $z \gtrapprox 0.05$, much more compact than has been previously assumed. Indeed, some these bright, knotty peaks of cosmic web emission are of an angular size that they could be easily confused with more compact emission sources, and indeed may already be present in the current generation of sky surveys.

Our simulation gives hope to the future detection of the cosmic web, primarily since the cosmic web emission and the extragalactic radio sources do not cluster in the same way. Our cross-correlation analysis shows: the cosmic web emission is clearly detectable both at 150 and \SI{900}{\mega \hertz} by way of cross-correlation, even in the presence of the much more luminous AGN and SFG sources; and that these embedded sources show negligible correlation, and in many cases, weak anti-correlation, with the cosmic web emission. In this case, the kernel used in the cross-correlation was the known cosmic web map itself; in a real detection attempt, a kernel will need to be devised with only approximate knowledge of the distribution of cosmic web flux. We hope the present simulation can aid in the development of such a kernel.

Besides the AGN and SFG populations, there are additional confounding factors that may impact detection techniques. One such factor in particular is the radio halo population which consists of large, extended low surface brightness features that are centrally located in cluster and group cores, and which without careful consideration could contribute to false or exaggerated signals. The introduction of this population into FIGARO could prove a useful direction in future work.

\section*{Acknowledgements}
F.V. acknowledges financial support from the ERC  Starting Grant "MAGCOW", no. 714196. The cosmological simulations on which this work is based have been produced using the ENZO code (http://enzo-project.org), running on Piz Daint supercomputer at CSCS-ETHZ (Lugano, Switzerland) under project s805 (with F.V. as PI, and the collaboration of C. Gheller and  M. Br\"{u}ggen).  We also acknowledge the usage of online storage tools kindly provided by the INAF Astronomical Archive (IA2) initiative (http://www.ia2.inaf.it).




\bibliographystyle{pasa-mnras}
\bibliography{main} 




\appendix

\section{Two-point correlation function}
\label{appendix:twopoint}

To compare clustering properties, we make use of the two-point correlation function $\xi\left(r\right)$. For a given set of points in a volume, this function estimates the probability of finding two points separated by some distance $r$ with respect to a set of points that were randomly (uniformly) distributed. We have calculated this function using the estimator provided by \citet{Landy1993}:
\begin{equation}
    \xi\left(r\right) = \frac{ \langle D,D \rangle -2f \langle D,R \rangle + f^2 \langle R,R \rangle }{f^2 \langle R,R \rangle}
    \label{eqn:2point}
\end{equation}
In this equation the notation $\langle A,B \rangle$ indicates the number of unique pairs of points ($a, b$), where $a \in A$ and $b \in B$, for which the condition $r < |a - b| < r + dr$ is satisfied. The set $D$ is our data, and contains the set of points for which we wish to calculate the two-point correlation function; the set $R$ is a set of points randomly assigned to the same volume according to a uniform distribution; $f = \nicefrac{ \text{length}\left(D\right) }{ \text{length}\left(R\right) }$ is a normalisation constant to account for any differences in the number of points in each set. To calculate the error in this estimate, we use the `bootstrap resampling' method (see, e.g. \citealp{Ling1986}) whereby we repeat the calculation for a number of independently generated $R$, over which we can calculate the mean and standard deviation of the resulting values.

In the case of the angular two-point correlation function, the Euclidean distance is replaced by the apparent angular distance in projection on the celestial sphere.

\section{Confusion Limit Calculation}
\label{appendix:confusion}

\begin{figure}
    \centering
    \includegraphics[width=\linewidth]{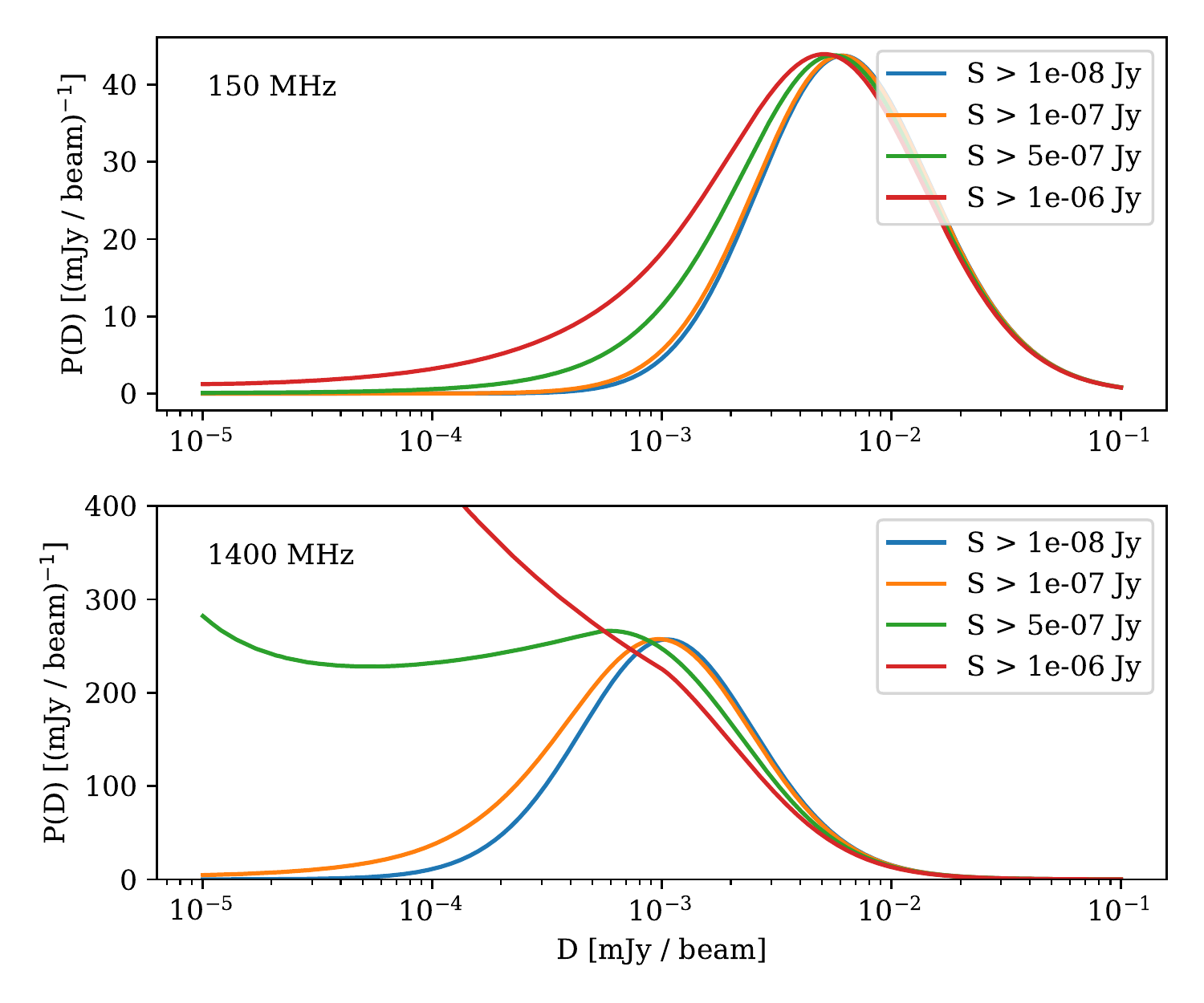}
    \caption{The probability of deflection for the T-RECS catalogue at \SI{150}{\mega \hertz} and \SI{1400}{\mega \hertz} with an \SI{8}{\arcsecond} circular Gaussian beam. Each curve shows the distribution for a different lower \SI{1400}{\mega \hertz} threshold cutoffs, showing that simulating sources down to \SI{0.1}{\micro \jansky} at \SI{1400}{\mega \hertz} is sufficient to simulate the classical confusion noise across this frequency range.}
    \label{fig:confusion}
\end{figure}

The choice of lower flux threshold when generating the T-RECS catalogue was selected so as to fully simulate classical confusion noise in the frequency range \SIrange[range-phrase=--,range-units=single]{150}{1400}{\mega \hertz} assuming a maximum resolution of \SI{8}{\arcsecond}. We make use of the `probability of deflection' or $P(D)$ technique that is described in \citet{Vernstrom2014}, and references therein, to calculate the classical confusion noise. This technique allows for us to calculate, for any given point on the sky, the probability distribution of the flux density per beam based upon two inputs: a differential source count ($\nicefrac{dN}{dS}$) and a synthesised beam shape. The classical confusion noise can then be estimated from the width of the peak in this distribution. In this case, we calculated the differential source count from the T-RECS catalogue itself, calculated independently at \SI{150}{\mega \hertz} and \SI{1400}{\mega \hertz}, and for the synthetic beam shape we assumed a circular Gaussian with FWHM of \SI{8}{\arcsecond}. In \autoref{fig:confusion} we compare the $P(D)$ distribution that results from a range of lower flux thresholds, that is to say, by setting $\frac{dN}{dS_{1400}}(S < S_0) = 0$ for a range of $S_0$. At \SI{150}{\mega \hertz}, for all the lower flux thresholds considered, we can see the central peak is well formed indicating that the confusion noise (at $\sim$\SI{18}{\micro \jansky \per \beam}) is well simulated. At \SI{1400}{\mega \hertz}, on the other hand, we can see that for the lower flux thresholds of \SI{5E-7}{\jansky} and \SI{E-6}{\jansky} the peaks are malformed, and only at \SI{E-7}{\jansky} and below is the confusion noise (at $\sim$\SI{3}{\micro \jansky \per \beam}) well simulated.

\section{Radial cross-correlation}
\label{appendix:crosscorrelation}

The radial cross-correlation of discrete maps $A$ and $B$ is constructed by first calculating its 2D cross-correlation function, defined as:
\begin{equation}
    R(\Delta x, \Delta y) = \sum_{i, j} \frac{ \left( A(i, j) - \bar{A} \right)\left( B(i + \Delta x, j + \Delta y) - \bar{B} \right) }{ N(\Delta x, \Delta y) \cdot \sigma_A \sigma_B }
\end{equation}
where $A(i, j)$ is the $(i,j)$th component of map $A$, $\bar{A}$ is the map mean, $\sigma_A$ is the standard deviation across the map, and $N(\Delta x, \Delta y)$ is a normalisation function. In essence, the maps are offset from each other by $\Delta x, \Delta y$ and we sum the product of all overlapping values, where the normalisation function simply counts the number of overlapping cells. The \textit{radial} auto-correlation function is simply the radial average of this function, with $r = \sqrt{\Delta x^2 + \Delta y^2}$, and where in practice we discretise radial values into bins and average over the 2D values of the function that fall within the bin.


\label{lastpage}
\end{document}